\documentclass[prd,aps,showpacs,tightenlines,preprint,nofootinbib,floatfix]{revtex4-1}
\pdfoutput=1

\usepackage{epsfig}
\usepackage{graphicx}
\usepackage{multirow} 
\usepackage{color}
\usepackage[dvipsnames]{xcolor}
\usepackage{amsmath,amssymb}
\usepackage{rotating}
\usepackage[caption=false]{subfig}
\usepackage{york}
\usepackage{hyperref}

\newcommand{\lp}{\left(}
\newcommand{\rp}{\right)}
\newcommand{\nn}{\nonumber}
\newcommand{\be}{\begin{equation}}
\newcommand{\ee}{\end{equation}}
\newcommand{\bea}{\begin{eqnarray}}
\newcommand{\eea}{\end{eqnarray}}

\def\ion#1#2{#1\,{\sc #2}\relax}  

\begin{document}
\preprint{TUM-HEP 1141/18}
\preprint{DESY 18-081}

\title{Lyman-$\alpha$~forest constraints on interacting dark sectors}

\newcommand\york{Department of Physics and Astronomy, York University,\\Toronto, Ontario, M3J 1P3, Canada}
\newcommand\tum{Technische Universit\"at M\"unchen, Physik-Department, James-Franck-Str.~1, 85748 Garching, Germany}
\newcommand\desy{Deutsches Elektronen Synchrotron DESY, Notkestr.~85, 22607 Hamburg, Germany}

\author{Mathias~Garny$^1$}
\email{mathias.garny@tum.de}
\author{Thomas~Konstandin$^2$}
\email{thomas.konstandin@desy.de}
\author{Laura~Sagunski$^3$}
\email{sagunski@yorku.ca}
\author{Sean~Tulin$^3$}
\email{stulin@yorku.ca}

\affiliation{$^1$\tum\\$^2$\desy\\$^3$\york}

\date{\today}

\begin{abstract}

The Lyman-$\alpha$ forest is a valuable probe of
dark matter models featuring a scale-dependent suppression of the power spectrum as compared to $\Lambda$CDM. 
In this work, we present a new estimator of the Lyman-$\alpha$ flux power spectrum that does not rely on hydrodynamical simulations. 
Our framework is characterized by nuisance parameters that encapsulate the complex physics of the intergalactic medium and sensitivity to highly non-linear small-scale modes. 
After validating the approach based on high-resolution hydrodynamical simulations for $\Lambda$CDM,
we derive conservative constraints on interacting dark matter models from BOSS Lyman-$\alpha$ data on large scales,  
$k<0.02\;($km/s$)^{-1}$, with the relevant nuisance parameters left free in the model fit. The estimator yields lower bounds on the mass of cannibal dark matter, where freeze-out occurs through $3 \to 2$ annihilation, in the MeV range. 
Furthermore, we find that models of dark matter interacting with dark radiation, which have been argued to address the $H_0$ and $\sigma_8$ tensions, are compatible with BOSS Lyman-$\alpha$ data.

\end{abstract}


\maketitle

\section{Introduction \label{sec:intro}} 

The evidence for dark matter (DM) spans from small dwarf galaxies~\cite{Strigari:2008ib} to the large scale structure of the observable Universe~\cite{Ade:2015xua}.
On large scales, precision cosmological data have converged on the $\Lambda$CDM model, where cold dark matter (CDM) makes up about 84\% of the total matter density~\cite{Ade:2015xua}.
Weakly interacting massive particles (WIMPs) are one leading particle physics candidate for CDM and many experimental efforts are underway to discover them through their interactions with visible matter~\cite{Bertone:2004pz,Feng:2010gw,Arcadi:2017kky}. 
However, if the CDM paradigm is not consistent across all cosmological and astrophysical probes, a new theory of DM must arise.
In this way, the particle physics of DM may be tested purely gravitationally, independent of any assumptions for its couplings to visible matter.

By assumption, CDM behaves as a pressureless perfect fluid with no interactions aside from gravity. 
However, many DM models predict a richer set of possibilities.
DM may comprise various species of matter and radiation within a dark sector~\cite{Carlson:1992fn,Boehm:2000gq,Boehm:2003hm,Strassler:2006im,Pospelov:2007mp,ArkaniHamed:2008qn}, with interactions between them---analogous to the visible sector of baryons, electrons, photons, neutrinos, and their known forces.
These effects can impact the structure and evolution of DM across different scales. 
Moreover, it may be that only a fraction of DM is interacting, while the remainder behaves as normal CDM.

The motivation for dark sector physics has grown recently in light of several `problems' with $\Lambda$CDM across a range of scales. 
On large scales, there are tensions between different probes for the Hubble rate $H_0$ and normalization of the matter power spectrum $\sigma_8$.
Their most precise values are inferred from the temperature and polarization anisotropies of the cosmic microwave background (CMB) by the Planck satellite, assuming $\Lambda$CDM cosmology~\cite{Ade:2015xua}. 
On the other hand, direct measurements of $H_0$ from the local distance ladder~\cite{Riess:2016jrr} and strong-lensing time delays~\cite{Bonvin:2016crt} yield larger values by $2-3\sigma$.
Additionally, weak lensing~\cite{Heymans:2013fya,Hildebrandt:2016iqg,Joudaki:2017zdt} and Sunyaev-Zel'dovich cluster counts~\cite{Ade:2015fva} favor smaller values of $\sigma_8$ by $2-3\sigma$. The level of discrepancy, however, is sensitive to how the systematics are treated~\cite{Ade:2015fva,Joudaki:2016mvz,Pan:2018zha}.
Lastly, there are several long-standing issues with CDM on small scales (see recent reviews~\cite{Weinberg:2013aya,DelPopolo:2016emo,Tulin:2017ara}).
Observations of many systems find a DM mass deficit in their inner halos compared to CDM predictions~\cite{Moore:1994yx,Flores:1994gz}.
These systems include dwarf galaxies near the Milky Way~\cite{Walker:2011zu} and in the field~\cite{Oh:2010ea}, low surface brightness spiral galaxies~\cite{McGaugh:1998tq,deBlok:2001hbg}, and massive clusters~\cite{Newman:2012nw}.
Moreover, the number of satellites in the Local Group~\cite{Moore:1999nt,Klypin:1999uc} and their internal kinematics~\cite{BoylanKolchin:2011de} have also been cited as challenges for CDM.

Although these issues remain inconclusive, it is nonetheless tantalizing to consider explanations beyond the CDM paradigm. 
Several proposals consider DM interacting with dark radiation to relieve the tension in $\sigma_8$ and/or $H_0$~\cite{Buen-Abad:2015ova,Lesgourgues:2015wza,Ko:2016uft,Buen-Abad:2017gxg,Raveri:2017jto}. 
For example, DM may be charged under a non-Abelian gauge symmetry, which has a coupling so weak it never confines~\cite{Buen-Abad:2015ova}. 
DM interactions with an ideal fluid of dark gluons creates a drag force to suppress perturbations that enter the horizon before matter-radiation equality, while not affecting larger scales relevant for the CMB that enter later. 
A related idea, dubbed partially acoustic dark matter (PAcDM)~\cite{Chacko:2016kgg,Raveri:2017jto}, achieves a similar effect. 
In this model, a small fraction of DM is tightly coupled with dark radiation (the remainder is CDM), suppressing perturbations that are smaller than the dark sound horizon.

In another class of models, DM may `cannibalize' itself through $3\to 2$ annihilation~\cite{Carlson:1992fn,deLaix:1995vi,Farina:2016llk}.
If entropy is conserved in the dark sector, the kinetic energy released by cannibalization partially compensates for energy lost due to Hubble expansion, keeping DM `warmer' compared to WIMPs of the same mass.\footnote{Strongly interacting massive particles (SIMPs) are a related scenario based on $3 \to 2$ scattering~\cite{Hochberg:2014dra,Hochberg:2014kqa}. However, SIMPs are identical to CDM as far as structure formation since the SM and dark sector are maintained in kinetic equilibrium. The framework of elastically decoupled relics~\cite{Kuflik:2015isi} interpolates between the SIMP and cannibal limits somewhat by allowing for the two sectors to fall out of equilibrium around the freeze-out epoch.}
During cannibalization, the DM temperature cools as $T_{\rm dm} \propto 1/\log a$, where $a$ is the scale factor, compared to the usual $T_{\rm dm} \propto 1/a$ for WIMPs in kinetic equilibrium with the SM. 
After cannibalization has frozen out and the DM relic density is fixed, the temperature falls as $T_{\rm dm} \propto 1/a^2$, similar to WIMPs that have fallen out of kinetic equilibrium.
MeV-scale cannibal DM behaves similarly to keV-scale sterile neutrinos (the canonical warm DM candidate), erasing small scale structure due to free-streaming~\cite{Viel:2005qj}, but without the associated X-ray signatures from DM decay~\cite{Abazajian:2017tcc}. 
Partial cannibal scenarios have also been considered~\cite{Buen-Abad:2018mas}.
Particle physics candidates include dark glueballs~\cite{Faraggi:2000pv,Cline:2013zca,Boddy:2014yra,Soni:2016gzf,Forestell:2016qhc,Halverson:2016nfq} or pions~\cite{Hochberg:2014kqa} from a non-Abelian theory.

The models we have discussed so far also have a natural connection to small scale structure. 
If DM interacts with dark radiation, then it is certain to self-interact as well~\cite{Ackerman:mha,Feng:2009mn,Bringmann:2013vra,Ko:2014bka,Chu:2014lja,Buckley:2014hja}, as does cannibal DM~\cite{Carlson:1992fn}.
Elastic self-scattering between DM particles can reconcile halo density profiles with observations in galaxies and clusters~\cite{Kaplinghat:2015aga}.
Recently, simulations have been made to explore the connection between DM self-interactions and deviations in the matter power spectrum beyond CDM~\cite{Buckley:2014hja,Cyr-Racine:2015ihg,Vogelsberger:2015gpr}.

In this work, we derive new constraints on interacting DM from the Lyman-$\alpha$ forest flux power spectrum. 
These observations come from absorption lines in the spectra of distant quasars due to neutral hydrogen gas clouds, which trace the underlying DM density~\cite{1986Ap&SS.118..509I,1986MNRAS.218P..25R}.
The Lyman-$\alpha$ forest is sensitive to intermediate scales between those relevant for large-scale structure (LSS) and the $\sigma_8$ tension, on the one hand, and the scales relevant for the small-scale puzzles on the other hand. 
Therefore, Lyman-$\alpha$ observations are relevant for models addressing potential large- as well as small-scale issues of the CDM paradigm, and provide a probe for models featuring a scale-dependent suppression of the power spectrum.

It is nontrivial to compute the flux power spectrum from first principles for a given DM model. 
The usual strategy is to model the properties of the intergalactic medium and then perform hydrodynamical simulations in order to extract the flux power spectrum (see, e.g.,~\cite{Bolton:2016bfs}). 
In general, each choice of
cosmological parameters as well as a set of parameters characterizing reaction rates within the intergalactic medium
must be run separately in a grid of simulations~\cite{Palanque-Delabrouille:2014jca,Palanque-Delabrouille:2015pga}.
There also exist approaches based on an effective model description with parameters determined by fitting to simulations~\cite{Arinyo-i-Prats:2015vqa,Murgia:2017lwo}. 

Here we propose a new effective framework for Lyman-$\alpha$ constraints on DM models based on perturbation theory that does not require simulations. 
We focus on measurements on comparably large scales with wavevector $k<0.02\:($km/s$)^{-1}$ from the Baryon Oscillation Spectroscopic Survey (BOSS)~\cite{Palanque-Delabrouille:2013gaa}. 
While the strongest Lyman-$\alpha$ constraints (e.g., for warm DM) have come from smaller scales based on MIKE/HIRES observations~\cite{Viel:2013apy,Irsic:2017ixq}, we hope to take advantage of the fact that the BOSS data lie within the weakly non-linear regime for the relevant redshifts $z\simeq 2-5$. However, this naive expectation is hampered by two related effects.
First, the relation between the flux power spectrum and the underlying DM density and velocity fields depends on the complex dynamics of the intergalactic medium. 
Second, the measured power spectrum is integrated across the line-of-sight of observation and therefore is effectively sensitive also to smaller, highly non-linear scales. 
However, we argue that for scales $k<0.02\:($km/s$)^{-1}$ relevant for BOSS data, these uncertainties can, to a certain extent, be captured through two redshift-dependent nuisance parameters. 
The high quality of BOSS data allows us to determine these unknown parameters by fitting directly to observational data, when accepting a moderate loss in sensitivity. 
This is the strategy we are following here. (Note that also analyses based on simulations usually adopt a number of additional nuisance parameters determined in this way~\cite{Palanque-Delabrouille:2014jca,Palanque-Delabrouille:2015pga}.)

The remainder of this work is organized as follows. 
In Sec.\,\ref{sec:model}, we discuss the relation between the Lyman-$\alpha$ flux power spectrum and the dark matter power spectrum, and describe the effective model used in our analysis. 
Next, we briefly review the BOSS Lyman-$\alpha$ data set in Sec.\,\ref{sec:data}.
Our main results are presented in Sec.\,\ref{sec:num}. 
We show that our model yields an excellent fit to the BOSS data for $\Lambda$CDM and we obtain a limit on warm DM as a benchmark comparison to other studies.
We derive limits on models of cannibal DM and DM-dark radiation interactions. 
In particular, we assess the consistency of the BOSS data with interacting DM models of Ref.~\cite{Buen-Abad:2017gxg}, which alleviate the $\sigma_8$ tension, as well as the model of Ref.~\cite{Pan:2018zha}.
We conclude in Sec.\,\ref{sec:dis}. The appendix contains additional details on the computation of non-linear
corrections to the power spectrum.

\section{Lyman-$\alpha$~model \label{sec:model}} 

The theoretical prediction of the Lyman-$\alpha$ flux power spectrum depends on the dynamics and properties of the intergalactic medium at redshifts $z\gtrsim 2$, and requires dedicated numerical simulations, see e.g.~\cite{Bolton:2016bfs}.
This complicates attempts to use  Lyman-$\alpha$ data
for parameter estimation, which typically require a large number of sample points in high-dimensional parameter spaces. Apart from grid-based interpolation techniques \cite{Palanque-Delabrouille:2014jca,Murgia:2017lwo}, one possible strategy is to model the relevant features of the flux power spectrum in order to obtain an approximate, but efficient estimator, see e.g.~\cite{Arinyo-i-Prats:2015vqa}.
Here we follow along these lines, and in particular attempt to capture the effects of the intergalactic medium by introducing a number of nuisance parameters that are determined by fitting the theoretical model to data. This strategy is possible due to exquisite measurements of the flux power spectrum, see Sec.\,\ref{sec:data}. Non-linear clustering of the dark matter density and velocity fields falls within the weakly non-linear regime at the relevant redshifts $z\sim 3$ and scales $k\lesssim 0.02$(km/s)$^{-1}$. 
Nevertheless, due to the integration across the line-of-sight, also smaller fluctuations are relevant, on which perturbative techniques fail, and the complex physics of the intergalactic medium dominates the non-linear dynamics. As we will see, within the description adopted here, the impact of modes with $k\gtrsim 0.02$(km/s)$^{-1}$ can be encapsulated to leading order in a counter-term that is treated as a free nuisance parameter as well.

After briefly reviewing how the one-dimensional Lyman-$\alpha$ flux power spectrum along the line-of-sight is related to the three-dimensional density power spectrum in the linear approximation, we describe the fitting model used in our analysis, and the parameters that enter.
We use natural units with $c=\hbar=k_B = 1$. 
Wavenumbers are converted from velocity space to comoving momentum space using a factor ${\cal H}(z)=a(z) H(z)$ where $a$ denotes the scale-factor and $H$ the Hubble rate. For example, the scale where thermal motion becomes relevant is given by the thermal broadening parameter $k_s \simeq \sqrt{m_p/T}$, where $m_p$ is the proton mass, corresponding to $k_s = 0.11$ (km/s)$^{-1}$ for a typical temperature $T = 10000$\,K associated to the intergalactic medium, which \emph{today} converts into the momentum $k_s = 11$~h/Mpc. This can be compared to the much larger scales $k < 0.02$ (km/s)$^{-1}$ for which we require a prediction of the flux power spectrum.

We want to model the transmission fraction $F = \exp(-\tau)$, where $\tau$ is the optical depth for Lyman-$\alpha$ photons, in terms of the dark matter power spectrum. Quite generally, the fluctuations in the transmission fraction
\be
\delta_F = \frac{F}{\bar F} - 1 \, ,
\label{deltaF}
\ee
where $\bar F$ denotes the average transmission fraction, depend on the density contrast $\delta$ and the dimensionless gradient of the peculiar velocity $v_p$ along the line of sight
\be
\eta = - \frac{1}{aH} \frac{\partial v_p}{\partial x_p} \, ,
\ee
where $x_p$ is the comoving coordinate.
Assuming a linear relationship this relation reads 
\be
\delta_F = b_{F\delta} \, \delta + b_{F\eta} \, \eta \, .
\ee
The density contrast $\delta$ and the velocity gradient $\eta$ are actually more directly related to the optical depth $\tau$, with coefficients being translated according to $b_{Fx} = b_{\tau x} \, \log \bar F$.
In Zel'dovich approximation \cite{Zeldovich:1969sb}, the gradient of the velocity is related to the density contrast via~\cite{Bernardeau:2001qr}
$\eta = f \mu^2 \delta$, where $\mu$ is the angle between the line-of-sight and the momentum mode under consideration, $\mu = k_\parallel/k$, and $f = d\log D/d\log a$ is the growth rate, where $D$ denotes the usual growth factor. Here $k_\parallel$ is the projection of the wavevector along the line-of-sight. In this approximation, the three-dimensional flux power spectrum fulfills the relation
\be
P_F(k,k_\parallel,z) = b_{F\delta}^2 (1 + \beta \mu^2)^2 P_L(k,z) \, ,
\ee
where we introduced $\beta = b_{F\eta} f / b_{F\delta}$
and $P_L$ is the linear density power spectrum.
Furthermore, the relation between the transmission fraction and the density contrast depends on the 
ionization history and one finds in Zel'dovich approximation~\cite{Hui:1996fh} 
\be
\label{bias_Zel}
\beta = \frac{1}{2 - 0.7(\gamma-1)} \, , 
\ee
where the adiabatic index $\gamma$ of the intergalactic medium depends on its properties and the reionization history~\cite{Hui:1997dp}. 

Below the Jeans scale $k_J= a H /c_s$, baryonic density fluctuations cannot collapse -- unlike dark matter fluctuations. It is defined via the sound velocity 
\be
c_s^2 = \frac{T \gamma}{\mu_p m_p} \, .
\ee
Here, $\mu_p m_p$  is the mean particle mass in the intergalactic medium  ($\mu_p = 0.6$), and $T$ its temperature.
Accordingly, the bias function $b_{F\delta}^2$ contains an additional suppression factor $\exp(-(k/k_F)^2)$. The filtering scale $k_F$ is hereby a redshift space average of the Jeans scale~\cite{Gnedin:1997td}
\be
\frac{1}{k_F(t)^2} = \frac{1}{D(t)} 
\int_0^t dt^\prime \, \frac{a^2(t^\prime)}{k^2_J(t^\prime)} 
\left[\frac{d}{dt^\prime} \left( a(t^\prime)^2  \, \frac{d}{dt^\prime} D(t^\prime) \right) \right]
\int_{t^\prime}^t \frac{dt^{\prime\prime}}{a^2(t^{\prime\prime})} \, .
\ee
Besides, there are several effects that suppress the observed power along the line-of-sight $k_\parallel$. This includes thermal broadening~\cite{Hui:1997dp}, redshift space distortions due to peculiar velocities~\cite{Scoccimarro:2004tg} and the finite resolution of the experimental observation. These effects can be taken into account by another exponential suppression scale $\propto\exp(-k_\parallel^2/k_s^2)$~\cite{Hui:1997dp}. The dominant effect is hereby the thermal broadening $k_s \simeq \sqrt{m_p/T}$.

In our analysis, we do not rely on the linear relation $\eta = f \mu^2 \delta$, but use instead the auto-correlations of the density contrast and the velocity gradients as well as their cross-correlation spectrum, similar to the effective model of redshift-space distortions discussed in~\cite{Scoccimarro:2004tg}.
In addition, we introduce a generalized bias parameter $\beta$ (see below for the precise definition).
It is known from simulations that the result (\ref{bias_Zel}) obtained in Zel'dovich approximation is not very accurate for the reionized intergalactic medium~\cite{McDonald:2001fe}. In our model, we will hence fit $\beta$ to the data. We allow for a redshift dependence, generalizing the Zel'dovich analysis, and introduce two fitting parameters $\alpha_{\rm bias}$ and $\beta_{\rm bias}$ (we use $z_{\rm pivot} = 3.0$)
\begin{equation}
\beta = \alpha_{\rm bias} \, [a(z_{\rm pivot})/a(z)]^{\beta_{\rm bias}}\,.
\end{equation}
The last ingredient is the impact of the Si\,III absorption that imprints a visible modulation in the observed flux power spectrum. Following the literature~\cite{McDonald:2004eu, Palanque-Delabrouille:2013gaa},  we model this effect with  
a factor 
\be
\kappa_{\rm SiIII} = 1  + 2 \left(\frac{\rm f_{SiIII}}{1 - \bar F}\right) \cos(\Delta V  \, k_\parallel) + 
\lp \frac{\rm f_{SiIII}}{1 - \bar F} \rp^2 \, ,
\label{SiIII}
\ee
with the mean transmission fraction $\log \bar F  = \alpha_F \, [a(z_{\rm pivot})/a(z)]^{\beta_F}$ parameterized by $\alpha_F$ and $\beta_F$, and two more parameters $\Delta V$ and ${\rm f_{SiIII}}$.

\begin{figure}[t]
\begin{center}
\includegraphics[width=0.49\textwidth]{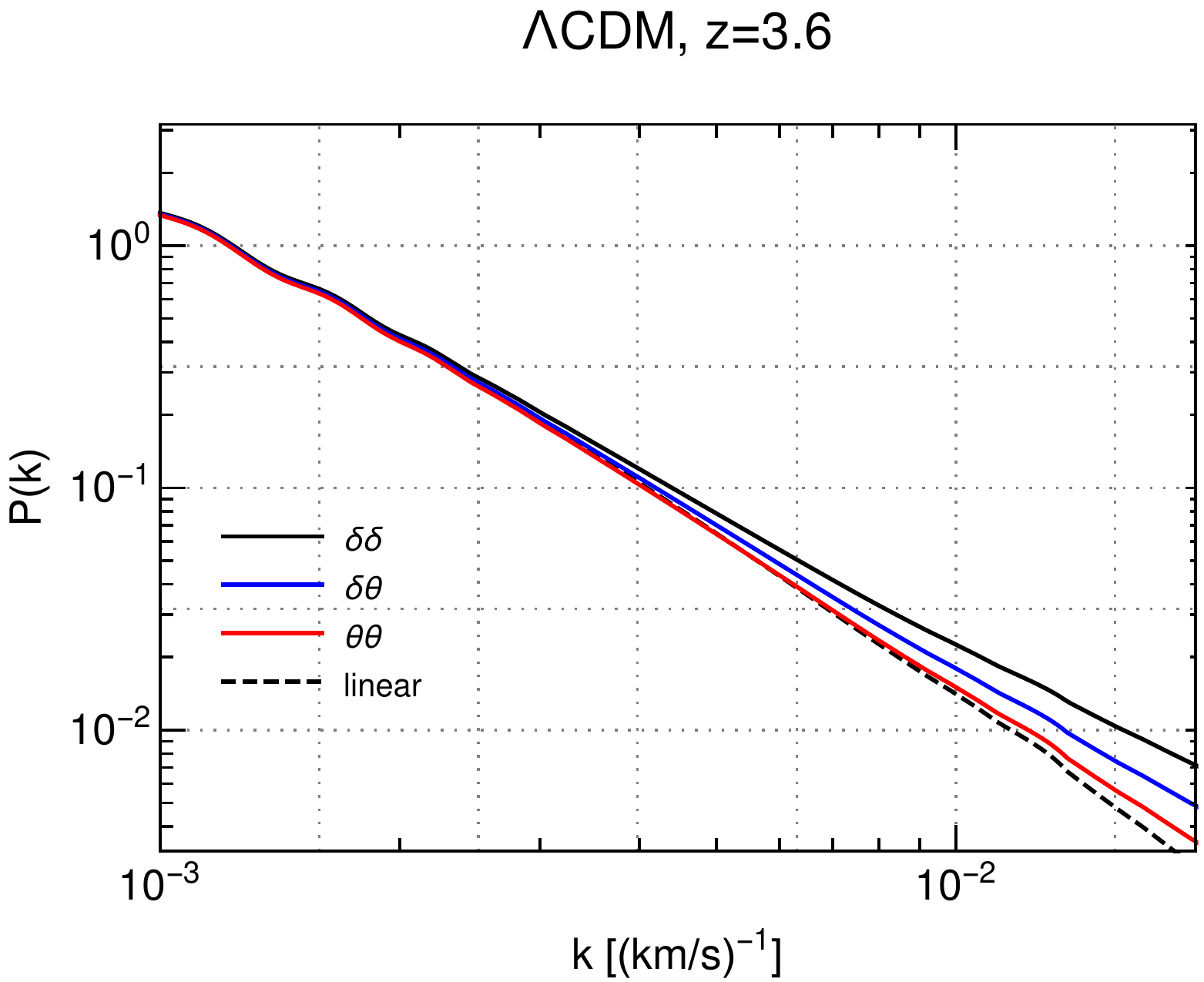}
\includegraphics[width=0.479\textwidth]{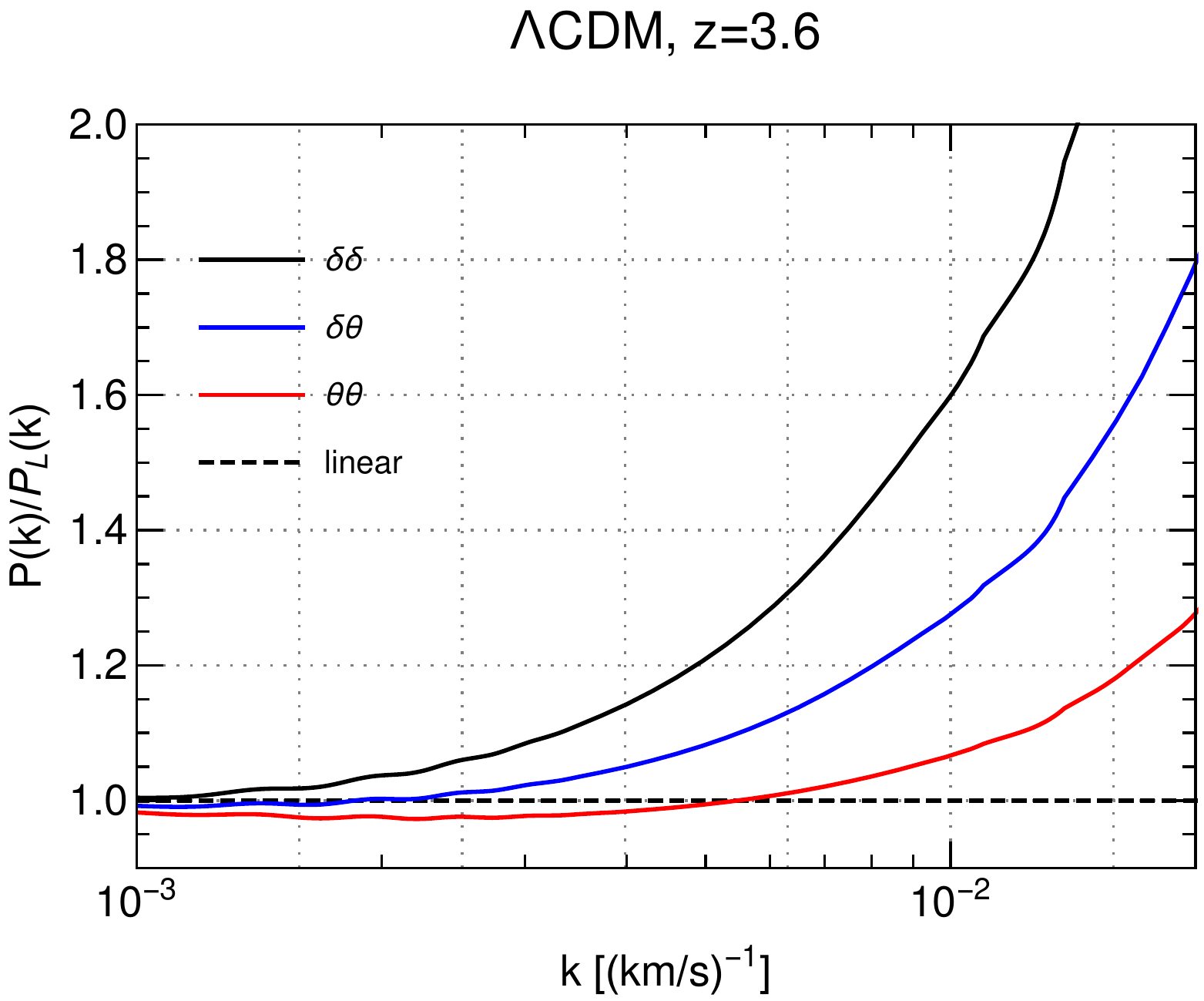}
\end{center}
\caption{\label{fig:P_deltadelta}%
\small Comparison of the non-linear two-loop power spectra $P_{\rm 2-loop}(k,z)$ of dark matter density-density ($\delta\delta$), density-velocity ($\delta\theta$) and velocity-velocity ($\theta\theta$) perturbations (left panel), normalized to the linear power spectrum $P_{L}(k,z)$ (right panel), for a $\Lambda$CDM Universe at redshift $z=3.6$. The linear power spectrum, which is identical for density, velocity and cross correlations, is indicated by the black dashed line.
}
\end{figure}

Ultimately, we are interested in the one-dimensional flux power spectrum~\cite{Kaiser:1990xe}, integrated across the line-of-sight,
\be
P_{\rm 1D}(k_\parallel,z) = \frac{1}{2\pi}\int_{k_\parallel} \, P_F(k,k_\parallel,z) \, k \, dk \, ,
\ee
and our model yields
\be\label{eq:model}
P_{\rm 1D}(k_\parallel,z) = 
A \, \kappa_{\rm SiIII}(k_\parallel,z) \, (\log \bar F(z))^2 \, 
\exp( - (k_\parallel/k_s(z))^2) \, ( I_0 + 2 \beta(z) I_2 + \beta(z)^2 I_4 ) \, , 
\ee
where we introduce a parameter $A$ for the overall amplitude, and
\bea
I_0(k_\parallel,z) &=& \int_{k_\parallel} \, dk \,k 
\, \exp(-(k/k_F)^2) \,  P_{\delta\delta}(k,z) + \bar I_0(z) \, , \nn \\
I_2(k_\parallel,z) &=& \int_{k_\parallel} \,  \frac{dk \,k_\parallel^2}{k} 
\, \exp(-(k/k_F)^2) \,  P_{\delta\theta}(k,z) \, , \nn \\
I_4(k_\parallel,z) &=& \int_{k_\parallel} \, \frac{dk \,k_\parallel^4}{k^3} 
\, \exp(-(k/k_F)^2) \,  P_{\theta\theta}(k,z) \, , 
\eea
where $P_{\delta\delta}(k,z)$ is the density power spectrum, $P_{\theta\theta}(k,z)$ the power spectrum of the velocity divergence $\theta=-\nabla\vec v/(aHf)$, and $P_{\delta\theta}(k)$ is the cross correlation. To obtain the non-linear power spectra, we first compute the initial power spectra with the Boltzmann code CLASS~\cite{2011arXiv1104.2932L}, and then determine their non-linear evolution by using the perturbative viscous fluid approach developed in~\cite{Blas:2015tla} at two-loop level (see App.\,\ref{sec:class} for details and Fig.~\ref{fig:P_deltadelta} for $\Lambda$CDM spectra).
For redshifts $z\sim 3$ and $k<0.02($km/s$)^{-1}$
the non-linear corrections are relevant, but well within the domain of validity of perturbation theory. In addition, we checked that the three-dimensional power spectra are insensitive to the UV cutoff used in the computation for $z\sim 3$.

However, there is one remaining issue concerning $I_0$, which depends on the power on short scales $k>0.02($km/s$)^{-1}$ in $P_{\delta\delta}$ (see Fig.~\ref{fig:P_deltadelta})
\footnote{%
$I_2$ and $I_4$ are dominated by lower wavenumbers below and up to the non-linear scale, for two reasons: (i) the factor $1/k$ or $1/k^3$ instead of $k$ in the integrand, and (ii) the smaller amplitude of $P_{\delta\theta}$ and $P_{\theta\theta}$ as compared to $P_{\delta\delta}$. Here the non-linear scale is defined as the wavenumber up to which the perturbative two-loop result for the density power spectrum is reliable, $k_{\rm nl}\simeq 0.03-0.05($km/s$)^{-1}$. The counterterm $\bar I_0$ may be considered as the leading correction in a systematic expansion in powers of $k_\parallel/k_{\rm UV}$, where $k_{\rm UV}={\rm min}(k_{\rm nl},k_s,k_F)$.}.
These scales are beyond the domain of validity of perturbative techniques and strongly influenced by non-linear physics. This introduces a cutoff dependence in the integral $I_0$, and we do not expect the model to fare particularly well in this regime. However, one point to note is that the difference $I_0(k_\parallel)-I_0(k_\parallel')$ for $k_\parallel,k_\parallel'<0.02($km/s$)^{-1}$ is insensitive to the problematic UV scales. In other words, the dependence of $I_0$ in the relevant range $k_\parallel<0.02($km/s$)^{-1}$ can be predicted up to a universal additive constant that captures the contribution from UV modes. In order to address this issue, we therefore introduce a counter-term (in the spirit of the effective field theory of large scale structure~\cite{Carrasco:2012cv}), $\bar I_0 = \alpha_{\rm c.t.} \, [a(z)/a(z_{\rm pivot})]^{\beta_{\rm c.t.}}$. We expect $\bar I_0$ to be due to non-linear clustering as well as non-linear dynamics of the intergalactic medium. For the former we expect a scaling similar to the leading non-linear correction, i.e. $\beta_{\rm c.t.} \gtrsim 4$, but impose no prior on this parameter. 
These additional parameters are introduced at the cost of predictivity. In turn, this means that statements on model exclusions are conservative. 
In total, our model thus includes eleven parameters,
\be
\{ A,\, \alpha_{F},\,  \beta_{F},\, \Delta V,\, f_{\rm SiIII},\, T,\, \gamma,\,  \alpha_{\rm bias},\, \beta_{\rm bias},\,\alpha_{\rm c.t.},\, \beta_{\rm c.t.} 
\}.
\ee

We also considered variations of the model by either reducing the number of free parameters (and making the model more predictive) or increasing the number of free parameters (and making the fit more accurate). For example, we added an overall nuisance parameter, $\exp(-\alpha k_\parallel^2)$, as often done in Lyman-$\alpha$ analyses using simulations~\cite{Palanque-Delabrouille:2014jca,Palanque-Delabrouille:2015pga}. This had the effect that suppression, for example by warm DM, is compensated by a negative nuisance parameter $\alpha$, which is unphysical. Restricting $\alpha$ to be positive lead to best-fit values close to zero and we discarded this parameter accordingly.

The ingredients described above seem to be all well motivated and indeed yield a very good fit to Lyman-$\alpha$ data (see Sec.\,\ref{sec:num}). From this standpoint there is no strong necessity to make the model more complicated, even if certain aspects of the intergalactic medium as well as possible contributions from non-linear bias are not resolved. One open question is whether it is justified to allow for a redshift dependence of the bias parameter $\beta$. On the one hand, this is for example seen in simulations~\cite{Arinyo-i-Prats:2015vqa}, on the other hand this is in disagreement with the prediction of the Zel'dovich approximation. We will present all numerical results for both cases and discuss the impact of this choice in the end.

In the following, we use our theoretical approach to predict the one-dimensional $\textnormal{Lyman-}\alpha$ flux power spectrum for different dark matter models. To constrain these models, we confront our theoretical predictions against Lyman-$\alpha$ measurements.

\section{Lyman-$\alpha$~Data \label{sec:data}} 

As a Lyman-$\alpha$ probe, we use the one-dimensional flux power spectrum~\cite{Palanque-Delabrouille:2013gaa} from the first release of quasar data of the BOSS~\cite{Dawson:2012va}, which is part 
of the Sloan Digital Sky Survey (SDSS-III) project~\cite{Eisenstein:2011sa,Gunn:2006tw,Smee:2012wd}.
The data consists of 13~821 quasar spectra 
selected from a larger sample of over 60~000 SDSS-III/BOSS DR9 spectra~\cite{Ahn:2012fh,Ross:2011ky}.
This tight selection of quasar spectra, based on high quality, high signal-to-noise ratio, and good spectra resolution, allowed to reduce the systematic uncertainties of the measurements to a lower level than the statistical uncertainties.

To determine the one-dimensional Lyman-$\alpha$ flux power spectrum from the selected sample of quasar spectra, two independent methods were applied in~\cite{Palanque-Delabrouille:2013gaa}. While the first method is based on a Fourier transform, the second one relies on a maximum likelihood estimator. 
We will refer to the one-dimensional Lyman-$\alpha$ flux power spectrum inferred with the Fourier transform method as `BOSS1 data', and to the one obtained with the likelihood method as `BOSS2 data'. 
The Lyman-$\alpha$ flux power spectra determined using either method are in good agreement over all redshift bins and wavenumbers. 
They are obtained in 12 redshift bins from $\langle z\rangle = 2.2$ to $\langle z\rangle = 4.4$, each bin spanning $\Delta z =0.2$, and 35 wavenumbers covering a range of $k_{\parallel}=0.001,\ldots,0.02\,$(km/s)$^{-1}$.

An example of the one-dimensional Lyman-$\alpha$ flux power spectrum for the `BOSS1 data' along with the best fit for CDM in the redshift range $\langle z\rangle = 2.6, \ldots, 3.6$ is shown in Fig.~\ref{fig:CDM}. 
The wiggles that appear in the Lyman-$\alpha$ flux power spectrum at all redshifts and with peak separations of $\Delta k = 2 \pi / \Delta V = 0.0028\,$(km/s)$^{-1}$ are due to the cross correlation between Lyman-$\alpha$ and \ion{Si}{iii} absorption (compare Sec.~\ref{sec:model}).

For the highest redshift bins  $\langle z\rangle  \geq 3.8$, the statistical uncertainty increases considerably, e.g. for $\langle z\rangle = 4.2$ and $\langle z\rangle = 4.4$, it exceeds $10\%$ on all scales. 
In addition to that, our numerical analysis indicates that the data points for the lower two redshift bins $\langle z\rangle = 2.2$ and $\langle z\rangle = 2.4$ do not contribute much to the fit because of their relatively large error bars, with a sizable contribution
from systematic uncertainties. Due to this, the final numerical results we present in the next section rely on an analysis which is restricted to the six redshift bins ranging from $\langle z\rangle = 2.6$ to $\langle z\rangle = 3.6$. Arguably, this choice also minimized the sensitivity to baryon feedback (becoming more and more important at low redshift $z\lesssim 3$) as well as uncertainties related to reionization (most relevant towards high redshifts $z\gtrsim 4$)~\cite{Bolton:2016bfs}.
Ultimately, the theoretical model for the Lyman-$\alpha$ flux power spectrum should be applied to the full likelihood function used by the BOSS collaboration. In this work, as a first step, we employ a simplified Gaussian treatment of observational uncertainties based on the published results, and adding the quoted statistical and systematic uncertainties in quadrature, in order to demonstrate the potential of the theoretical model description.

\section{Numerical results \label{sec:num}} 

In this section we apply the fitting procedure discussed in Sec.\,\ref{sec:model} to the BOSS Lyman-$\alpha$ data
described previously and test the compatibility with a number of benchmark models for self-interacting dark matter. Before that, we briefly discuss the case of cold and warm dark matter, as well as checks based on simulated mock data. 

Some of the parameters entering Eq.\,\eqref{eq:model} for the flux power spectrum only have a minor impact on the final outcome. This includes $\alpha_F$ that is for most parts degenerate with the overall amplitude parameter $A$ (and besides only marginally changes $\kappa_{\rm SiIII}$). We therefore fix the corresponding parameters to reasonable values inferred either by simulations~\cite{Arinyo-i-Prats:2015vqa} or observations~\cite{Palanque-Delabrouille:2013gaa}. These parameters are summarized in Tab.~\ref{tab:fixed}. 
Note that for the parameterization used in this work the temperature $T$ as well as the adiabatic index $\gamma$ enter explicitly only via the suppression scales $k_s$ and $k_F$, that are marginally relevant on the scales measured by BOSS. Since for example the counter-term and the velocity bias are left free in the fit, their implicit dependence on the properties of the intergalactic medium is not affected by the choice in Tab.~\ref{tab:fixed}.

\begin{table}
\begin{center}
  \begin{tabular}{ | c | c | }
    \hline
    $\alpha_F$ & $0.446$  \\ 
    \hline
    ${\rm f_{SiIII}}$ & $0.0081133$  \\ 
    \hline
    $\Delta V$  [km/s] & $2271.0$  \\
    \hline
    $T$  [K] & $10000$  \\
    \hline
    $\gamma$  & $1.0$  \\
    \hline
  \end{tabular}
\caption{\label{tab:fixed}%
All parameters that are fixed in the fit.
\small 
}
\end{center}
\end{table}
\begin{figure}[t]
\begin{center}
\includegraphics[width=0.48\textwidth]{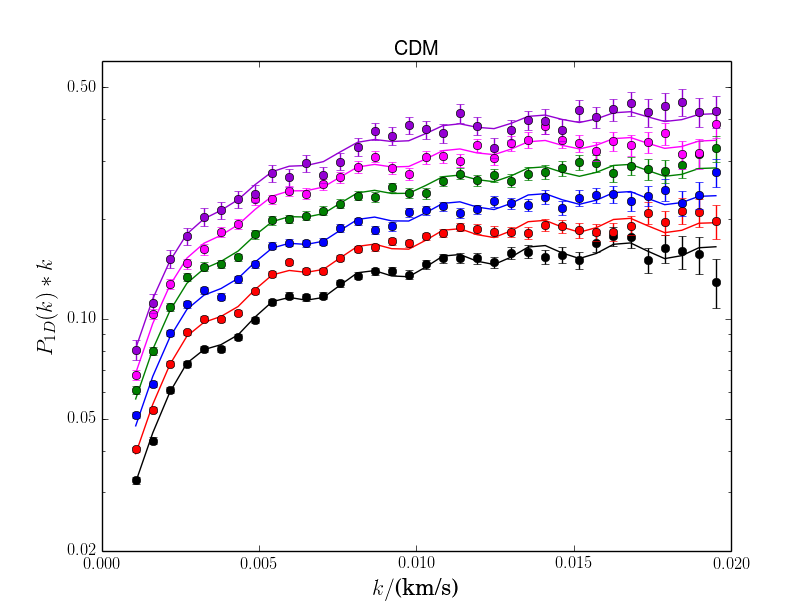}
\includegraphics[width=0.48\textwidth]{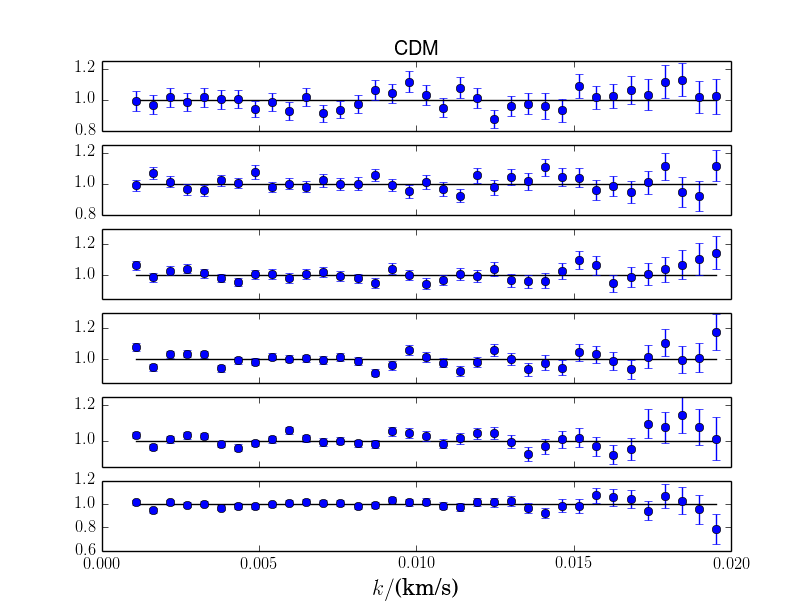}

\includegraphics[width=0.48\textwidth]{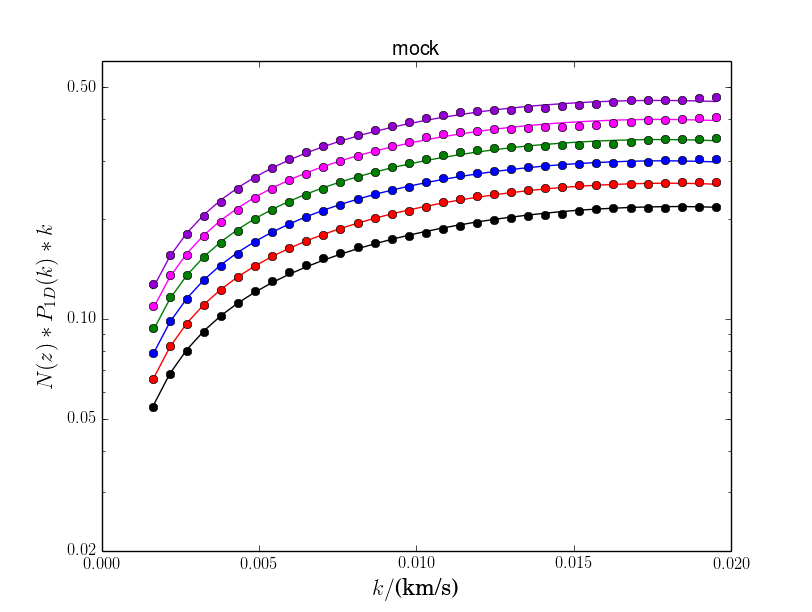}
\includegraphics[width=0.48\textwidth]{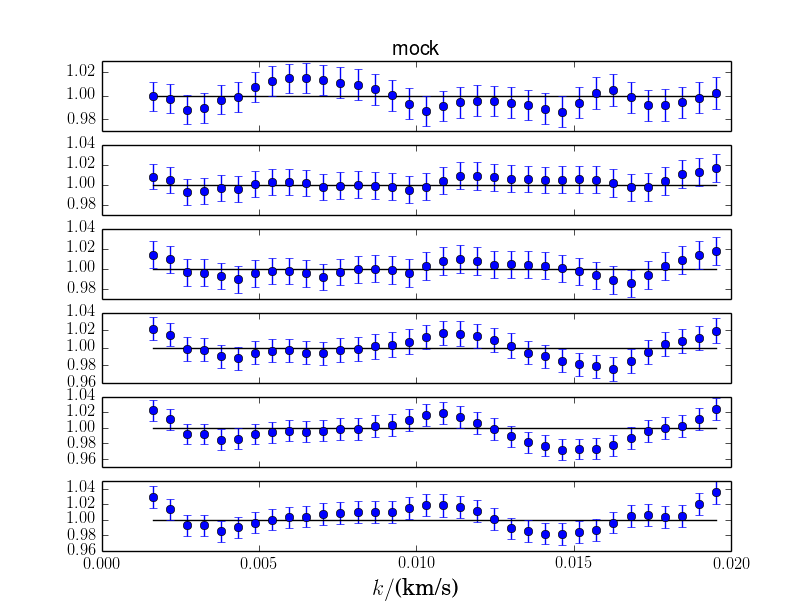}

\includegraphics[width=0.48\textwidth]{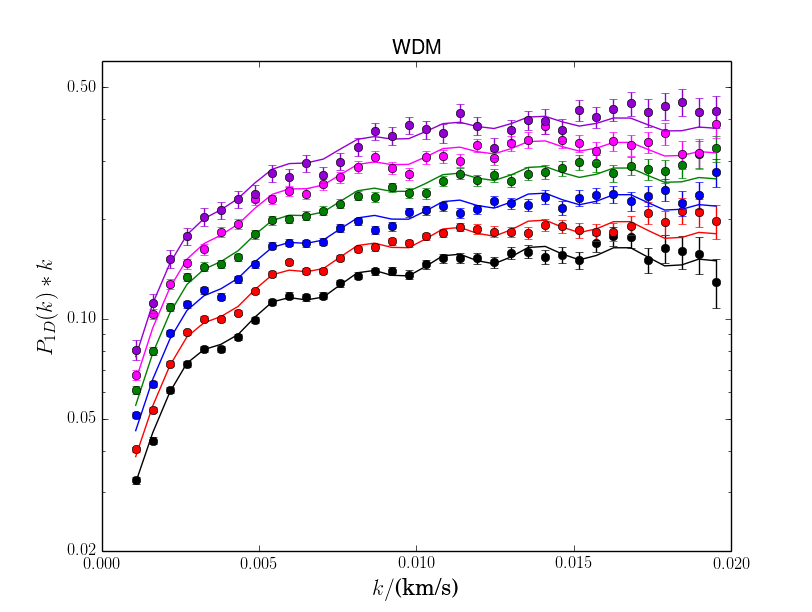}
\includegraphics[width=0.48\textwidth]{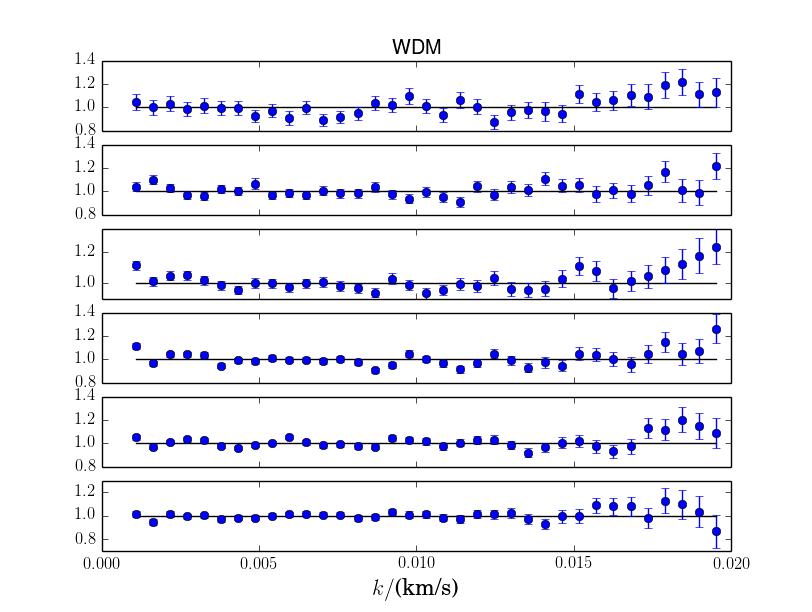}
\end{center}
\caption{\label{fig:CDM}%
\small \emph{Upper row:} Best fit of the Lyman-$\alpha$ flux power spectrum for a $\Lambda$CDM Universe to the BOSS1 data. On the left, we show our theoretical model prediction for the Lyman-$\alpha$ flux power spectrum $P_{1D}(k,z)$ (solid lines) as a function of the wavevector along the line-of-sight, $k$, and for redshifts $z=[2.6,2.8,3.0,3.2,3.4,3.6]$ (from bottom to top) in comparison to the BOSS1 data (dots). The corresponding residuals of the model fit for the same redshifts are plotted on the right.
\emph{Middle row:} Comparison of the Lyman-$\alpha$ flux power spectrum to mock data extracted from high resolution 
hydrodynamical simulations~\cite{Bolton:2016bfs} (see text).
\emph{Lower row:} As first row, but for a warm dark matter model with $m_{{\rm sterile}\ \nu}=1.6$keV, corresponding to $m_{\rm thermal}=0.46$keV. Note the different $y$-ranges in the right panel.
}
\end{figure}

Figure~\ref{fig:CDM} (upper row) shows the best fit of BOSS1 data to the $\Lambda$CDM model corresponding to Planck 2015~\cite{Ade:2015xua} cosmological parameters (see Tab.~\ref{tab:benchmark_param}). The corresponding best-fit parameters for the flux power spectrum are given in Tab.\,\ref{tab:fit}. 
The resulting $\chi^2$ of the fit with arbitrary $\beta_{\rm bias}$ is $231.7$ for $210$ data points (compare Tab.~\ref{tab:final1}). 
We checked that the result is stable when including the parameters from Tab.~\ref{tab:fixed} in the fit, or allowing for a more general parameterization, e.g. with redshift-dependent temperature with a single or broken power law. 

\vskip 6pt
In order to check the validity of the procedure outlined above, we compare to mock Lyman-$\alpha$ flux power spectra extracted from data of the Sherwood Simulation Suite~\cite{Bolton:2016bfs,SherwoodSimulation}. 
The Sherwood simulations are high-resolution state-of-the-art hydrodynamical simulations of the intergalactic medium for a $\Lambda$CDM cosmology, performed with a modified version of the GADGET-III code. 

The mock Lyman-$\alpha$ flux power spectra we use to validate our fitting procedure are drawn from data sets of high-resolution Sherwood simulations with $2048^3$ particles in a comoving volume of 
$40^3\,$h$^{-1}\,$Mpc$^{-3}$
at redshifts $2\leq z \leq 7$ ($\Delta z =2$). Each data set contains the Lyman-$\alpha$ optical depths $\tau$ for 2048 pixels along 5000 different lines-of-sight at a single redshift. From the optical depths, we compute the Fourier transform of the fluctuations in transmission fraction $\delta_{F}$ (see~\eqref{deltaF}). To obtain the one-dimensional Lyman-$\alpha$ flux power spectrum, we then determine the two-point correlation function and average over the line-of-sights such that $P_{\rm 1D} \sim \langle | \delta_{F}|^2 \rangle$ (see, e.g.,~\cite{Meiksin:2007rz} for details). 
The uncertainties in our mock dataset include statistical errors on the mean from averaging over lines-of-sight. We do not include systematic errors; for example, modeling uncertainties in the reionization history and treatment of baryonic feedback would increase the error budget by $5-15\%$ on the relevant scales and redshifts~\cite{Bolton:2016bfs}.

Using redshifts and $k$-values identical to those of the BOSS1 data (apart from the lowest $k$-bin), we apply the same analysis pipeline for the mock simulation data as for the BOSS data, except from the normalization of the mean flux, that is left free, and the Si~III modulations, that are not included in the simulation data. 
Figure~\ref{fig:CDM} (middle row) shows the
comparison of our fitted model for the one-dimensional flux power spectra (solid lines) to the mock data points~\cite{Bolton:2016bfs} (left panel), as well as the residuals and error bars corresponding to the statistical uncertainty of the mock spectra (right panel).
We find that our model~\eqref{eq:model} yields an excellent fit within the statistical uncertainty, nominally giving $\chi^2=126$ for $204$ degrees of freedom.\footnote{We treat the uncertainties of the mock data as being uncorrelated. One may question this assumption given that our fit to them is ``too good'' and the evident correlation between data points of the simulated spectra in Fig.~\ref{fig:CDM} (middle, right).}
The corresponding best-fit model parameters are shown in the last column of Tab.\,\ref{tab:fit}, and are comparable in magnitude to those obtained when using BOSS data.
Note that the parameter $\beta_{\rm bias}$ has opposite sign. This can be traced back to a slight difference between the BOSS and mock data for low $z$ and towards high $k$ values.
The difference is compatible with the expected
accuracy of the hydrodynamical simulation, as discussed above, and the systematic uncertainties quoted by the BOSS collaboration~\cite{Palanque-Delabrouille:2013gaa}.

\subsection{Warm dark matter \label{sec:warm}} 

The main purpose of this section is to benchmark the effective model for the flux power spectrum described in section~\ref{sec:model} and compare the limit on the warm DM mass to the ones obtained from full-fledged hydrodynamic simulations. 
The free streaming of warm DM leads to a suppression of the power spectrum at small scales, with a suppression scale that depends on the production mechanism. For production mechanisms that yield a spectrum $f=\chi/(e^{p/T_x}+1)$, with an arbitrary normalization factor $\chi$,
the initial linear power spectrum can be characterized by the temperature $T_x$, the dark matter mass $m_x$ and the density parameter $\omega_x=\Omega_x h^2$, given by \cite{Viel:2005qj} 
\begin{equation}
 \omega_x = \chi \left(\frac{T_x}{T_\nu}\right)^3\frac{m_x}{94{\rm eV}}\;.
\end{equation}
Two common assumptions are either $\chi=1, T_x\ll T_\nu$, corresponding to a fermionic particle that was in equilibrium with the thermal bath of the visible sector and decoupled at temperature $T_D\gg m_x$ when $g_*(T_D)=10.75(T_\nu/T_D)^3$, or $T_x=T_\nu, \chi\ll 1$ for a sterile neutrino produced via a small mixing with the active neutrinos.
In the following we assume that $\omega_x\simeq 0.12$
corresponds to the observed value \cite{Ade:2015xua},
which fixes $T_x$ for a thermal relic and $\chi$ for a
sterile neutrino, respectively, for any given mass $m_x$.

The power spectrum depends only on the ratio $T_x/m_x$.
Therefore, production mechanisms that lead to a different temperature can be related to each other by rescaling the dark matter mass. For example, the mass of a sterile neutrino that is produced with a temperature identical to the one for active neutrinos is related to an early-decoupled thermal relic by \cite{Viel:2005qj}
\begin{equation}
m_{{\rm sterile}\ \nu} \simeq 4.47 {\rm keV}\left(\frac{m_{{\rm thermal}}}{\rm keV}\right)^{4/3}\left(\frac{0.12}{\omega_x}\right)^{1/3}\;.
\end{equation}

\begin{figure}[t]
\begin{center}
\includegraphics[width=0.49\textwidth]{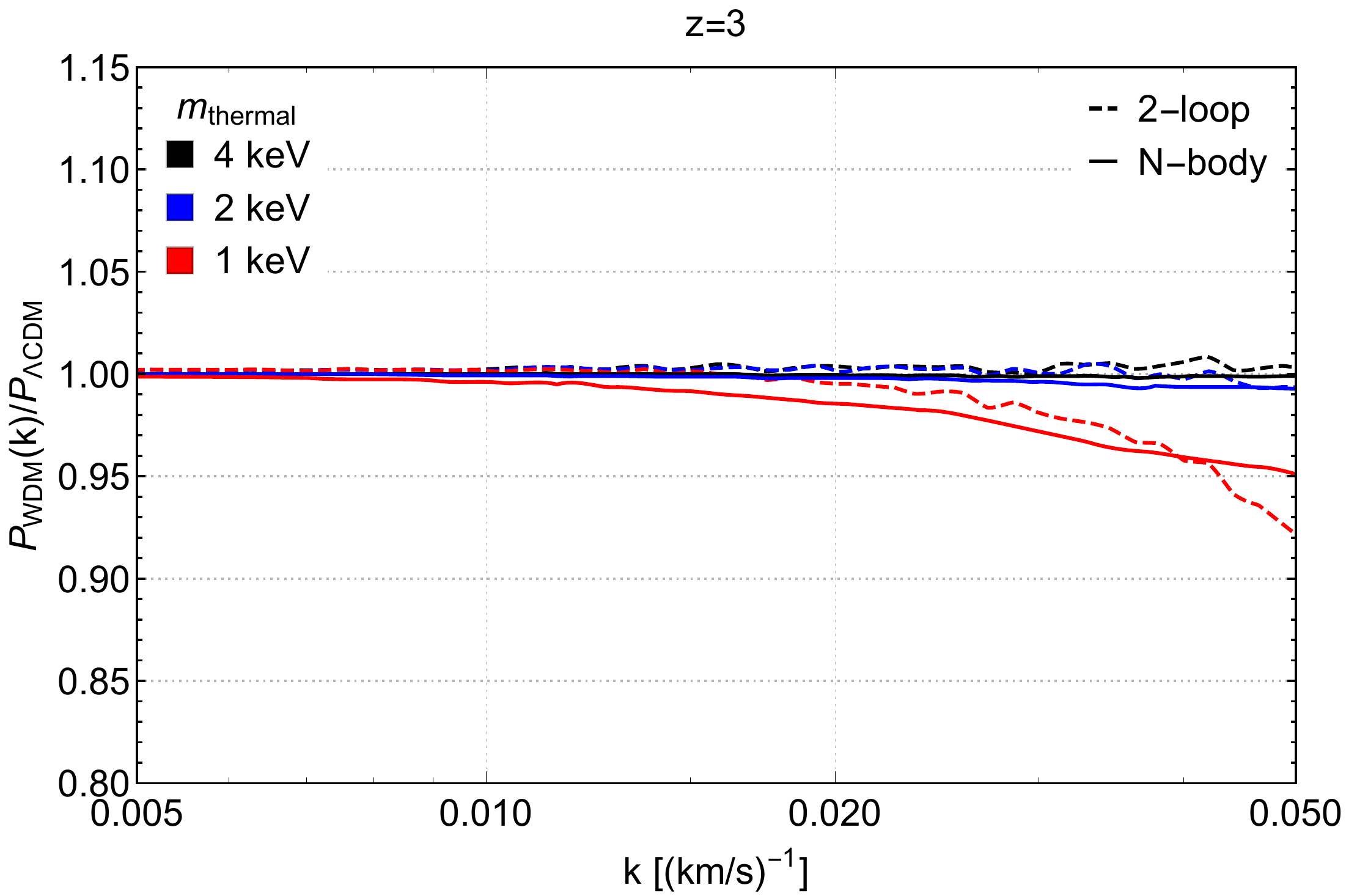}
\includegraphics[width=0.49\textwidth]{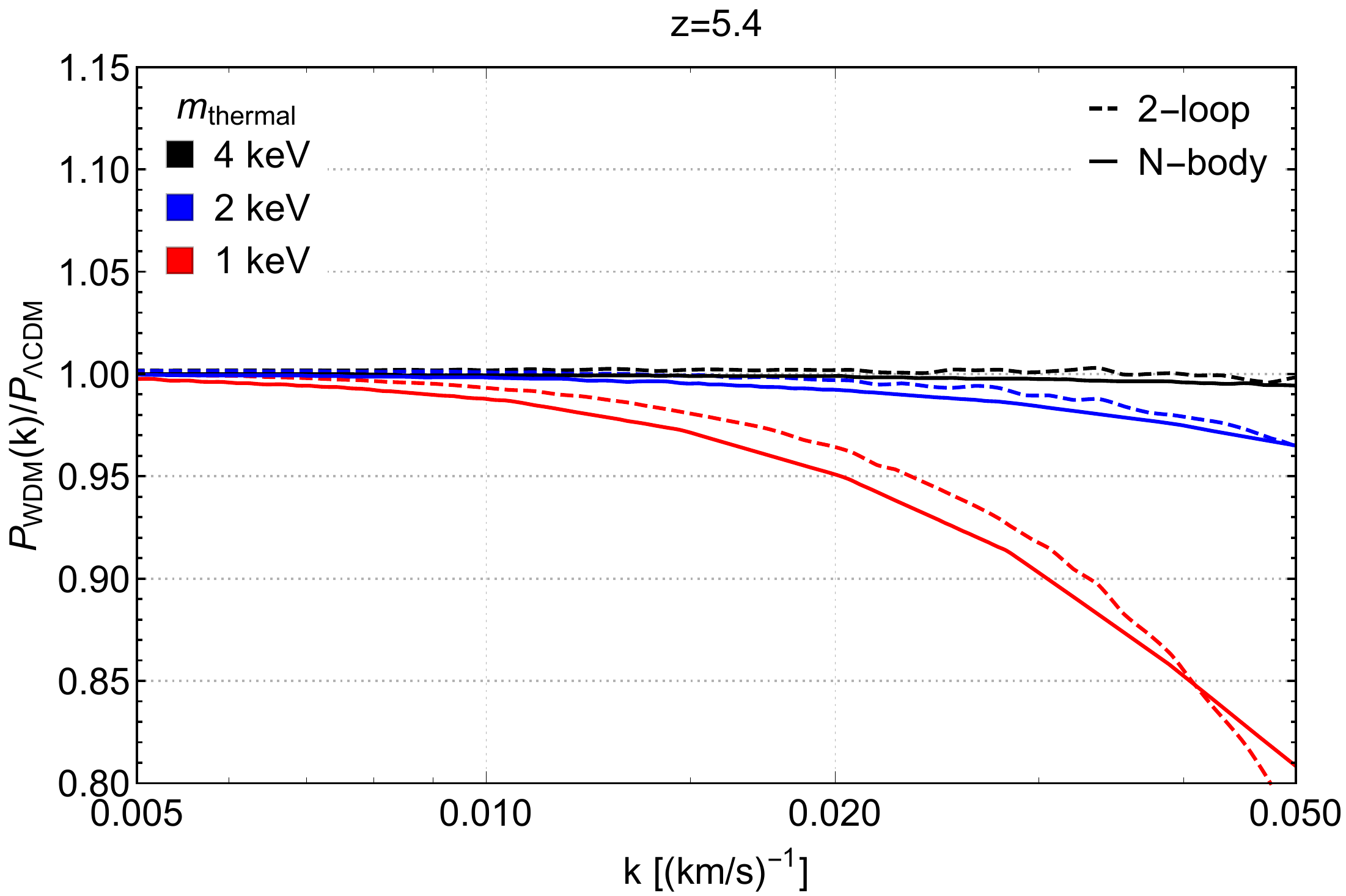}
\end{center}
\caption{\label{fig:WDM}%
\small Ratio between the non-linear density power spectrum for warm and cold dark matter, $P_{\rm WDM}(k)/P_{\rm \Lambda CDM}(k)$, at redshifts $z=3$ (left panel) and $z=5.4$ (right panel) for three different thermal WDM masses ($m_{\rm thermal}=4,2,1\,$keV in black, blue, red). We show the perturbative 2-loop predictions based on the viscous fluid approach~\cite{Blas:2015tla} (dashed lines) in comparison to the results from hydrodynamical simulations~\cite{Viel:2013apy} (solid lines). The initial density power spectra were generated with the CLASS code~\cite{2011arXiv1104.2932L}.
}
\end{figure}

We show the ratio of non-linear 3D warm- and cold dark matter power spectra for three masses $m_{\rm thermal}=1,2,4$\,keV in Fig.~\ref{fig:WDM}. Notably, the suppression of the power spectrum is more pronounced at early redshifts (right panel). At lower redshift non-linear evolution partially compensates the power suppression due to mode coupling (left panel). For comparison, we also show results from numerical simulations \cite{Viel:2013apy} based on the GADGET-II code \cite{Springel:2005mi}.
Since our results show reasonable agreement with these simulations, we are justified in our perturbative approach for the input power spectra in our model.

\begin{figure}[t]
\begin{center}
\includegraphics[width=0.7\textwidth]{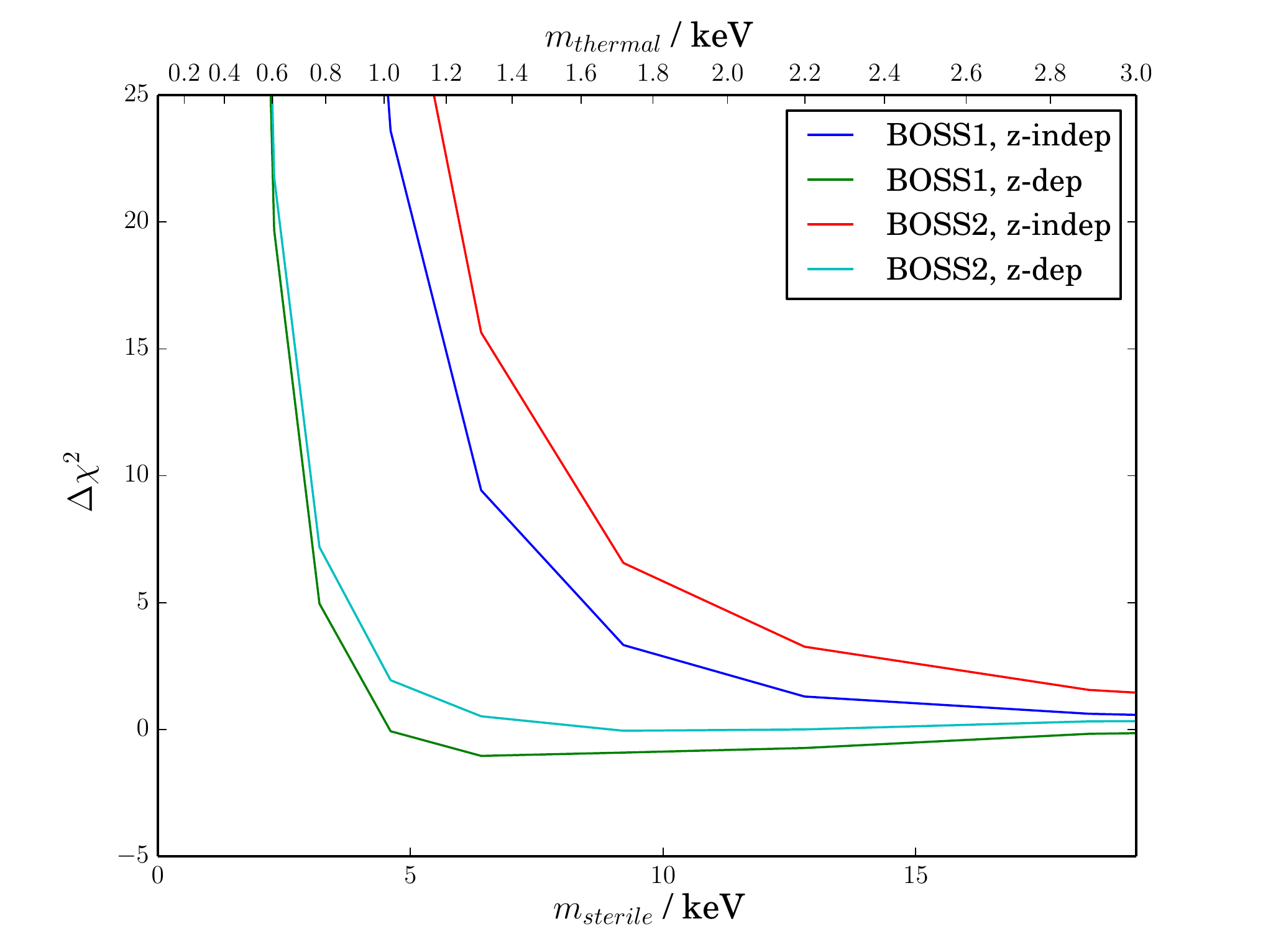}
\end{center}
\caption{\label{fig:boss_wdm}%
\small Best fit values of $\Delta\chi^2$ relative to $\Lambda$CDM, versus warm dark matter mass for the BOSS1 and BOSS2 data~\cite{Palanque-Delabrouille:2013gaa}. Using a redshift dependent bias $\beta$ (`$z$-dep') leads to  more conservative exclusion limits on the warm DM mass as compared to a constant bias (`$z$-indep'). Implementing a prior $\beta_{\rm bias} < 1$ for the redshift-dependence has no effect and the result coincides with the `$z$-dep' case.
}
\end{figure}

Our results are summarized in Fig.~\ref{fig:boss_wdm}. We show the value of $\Delta \chi^2$ relative to the cold dark matter case. 
One can see that using the model with a redshift-dependent bias $\beta$, the resulting limit is weaker than in the model without redshift-dependent bias. To be conservative, we will hence mostly focus on the model with a redshift-dependent bias when interpreting the data in models with interacting DM.
For illustration, we also include the Lyman-$\alpha$ flux power spectra for a warm dark matter model with very small mass $m_{{\rm sterile}\ \nu}=1.6$keV, corresponding to $m_{\rm thermal}=0.46$keV, in Fig.\,\ref{fig:CDM} (lower row), for which we find $\Delta\chi^2=70$ when using a redshift-dependent bias.

Ultimately, the limit we derive from the BOSS data~\cite{Palanque-Delabrouille:2013gaa} is of the order $m_{{\rm sterile}\, \nu}\gtrsim 10(5)$\,keV that translates into $m_{\rm thermal} > 2(1)$\,keV for redshift independent (dependent) bias. This bound is not quite competitive with the values in the literature\footnote{For example, the analysis of \cite{Baur:2015jsy} reports a limit $m_{\rm thermal} > 2.97$\,keV based on BOSS data~\cite{Palanque-Delabrouille:2013gaa} for $z\leq 4.1$.}, but it is encouraging that our simple analytic model has significant predictive power in this case. 
We also checked various modified versions by varying in addition parameters from Tab.\,\ref{tab:fixed}, introducing an additional overall suppression as described in Sec.\,\ref{sec:model}, or varying the UV cutoff used for the computation of the non-linear power spectra, and find that our results are stable. In addition, we checked that  the limit on the warm dark matter mass does not significantly depend on the other cosmological parameters within their range of uncertainty. A more restrictive limit could be derived when imposing a prior on the counter-term $\bar I_0$. However, in this work we prefer to follow a conservative approach and allow for arbitrary values. We therefore conclude that the procedure outlined above can be used to obtain rather conservative, albeit robust, constraints for any type of input power spectrum at relatively low computational cost.



\subsection{Strongly self-interacting dark matter}

Here we consider the class of models of strongly self-interacting dark matter, for which the dark matter momentum distribution is affected by cannibalization, i.e., number-changing self-annihilations.
In order to understand the impact on structure formation we need to consider the phase-space distribution. For thermal warm dark matter it is given by a redshifted, thermal distribution $f=1/(e^{p/T_{\rm WDM}}\pm 1)$. In contrast, we assume that the relevant momentum modes of strongly self-interacting dark matter remain in kinetic equilibrium among themselves until they are non-relativistic, due to efficient elastic self-interactions. The distribution function can then be taken to be $f\propto \exp\left(-\frac{p^2}{2mT(a)}\right)$, with $m$ being the dark matter mass and $T(a)$ the temperature. Its evolution depends on the type of interactions that are in equilibrium in a given epoch: For the case when \emph{elastic scatterings with the SM}  are in equilibrium, $T\propto T_{\rm SM}\propto g_{*S}^{-1/3}/a$. When this is not the case, but \emph{number-changing self-interactions} are efficient, $T\propto 1/\ln a$~\cite{Carlson:1992fn,deLaix:1995vi}. In general number-changing self-interactions will freeze out at some point, and subsequently temperature scales in the standard way of a decoupled non-relativistic species $T\propto 1/a^2$ until today. Note that the latter scaling applies irrespective of whether \emph{elastic self-scatterings} are efficient or not.

Dark matter with a non-relativistic momentum distribution has an equation of state $w=p/\rho=T/m$ and adiabatic sound velocity
\begin{equation}
c_s^2 = \frac{\dot p}{\dot \rho}=\frac{T}{m}\left(1-\frac13\frac{d\ln T}{d\ln a}\right)\,.
\end{equation}
In the non-relativistic limit $w\ll 1$ the main
impact on the evolution of the density contrast $\delta$ and velocity divergence $\theta$ can be captured by the sound velocity,
\begin{eqnarray}\label{eq:pert_sidm_nonrel}
  \dot\delta &=& -\theta +3\dot\phi\,, \nn\\
  \dot\theta &=& -{\cal H}\theta + c_s^2k^2\delta +k^2\psi \,,
\end{eqnarray}
where $\phi, \psi$ are the metric perturbations in conformal Newtonian gauge, ${\cal H}=aH$ is the conformal Hubble rate,
and contributions from higher moments are suppressed for $w\ll 1$~\cite{Blas:2014hya}.
For comparison, the sound velocity for fermionic warm dark matter, in the limit $T_{\rm WDM}\ll m_{\rm WDM}$, is given by \cite{Shoji:2010hm}
\begin{equation}
c_s^2|_{\rm WDM} = \frac{5\zeta(5)}{\zeta(3)}\left(\frac{T_{\rm WDM}}{m_{\rm WDM}}\right)^2\,,
\end{equation}
which scales as $1/a^2$ since $T_{\rm WDM}\propto 1/a$.

The redshift when the perturbation modes relevant for the BOSS Lyman-$\alpha$ data enter the horizon correspond to $z\lesssim 10^6$. In the following we assume that the temperature $T$ has already started to scale as $1/a^2$ by then, i.e. that the freeze-out of number-changing interactions occurs for some $z_f>10^6$. In this case $c_s^2\propto T/m$ scales exactly like the sound velocity for warm dark matter. Therefore, up to higher order corrections in $T/m$ and in $T_{\rm WDM}/m_{\rm WDM}$, the resulting power spectrum has approximately the same form as for warm dark matter. We checked that the linear matter power spectrum obtained from solving \eqref{eq:pert_sidm_nonrel} agrees with the warm dark matter power spectrum to better than $2\%(1\%)$ for $m_{\rm WDM}\equiv m_{\rm thermal}>1(2)$keV on the relevant scales $k\lesssim 5h/$Mpc, when choosing the normalization of $c_s$ appropriately.

The rescaling necessary to match the power spectra can be reproduced approximately by matching the sound velocity to the one of warm dark matter, which gives
\begin{equation}
\frac{T_0}{m}=r\times \frac35\times 10^{-14}\left(\frac{\rm keV}{m_{\rm WDM}}\right)^{8/3}\left(\frac{\omega_{\rm dm}}{0.12}\right)^{2/3}\,,
\end{equation}
where $T=T_0/a^2$ is the temperature of self-interacting dark matter after number-changing interactions have frozen out, and $T_0$ is its value rescaled to today. The direct matching of sound velocity gives $r=1$. Numerically, we find the best agreement of the linear power spectra when choosing $r=1.40,1.27,1.15$ for $m_{\rm WDM}=1,2,4$\,keV.
Therefore, the lower bound on the warm dark matter mass can be translated into an upper bound on $T_0/m$. 

Assuming the freeze-out of number-changing interactions takes place for a given ratio $x_f=m/T_f\gg 1$, the upper bound on $T_0/m$ can be interpreted as a lower bound on the redshift when the freeze-out occurs, $z_f>(x_f\times(T_0/m)_{\rm max})^{-1/2}$. This gives the lower bound
\begin{equation}
  m \geq 1.3\,{\rm MeV}\,\left(\frac{\xi_f}{100}\right)
  \left(\frac{x_f}{10}\right)^{\frac12}
  \left(\frac{g_{*S}(T_f)}{10.75}\right)^{\frac13}
  \left(\frac{m_{\rm WDM}^{\rm min}}{\rm keV}\right)^{\frac43}
  \left(\frac{0.1188}{\omega_{\rm dm}}\right)^{\frac13}
  r^{-\frac12}\,,
\end{equation}
where $\xi_f=T_f/T_{\rm SM}$ is the ratio of the self-interacting dark matter temperature at freeze-out to the Standard Model temperature, and $m_{\rm WDM}^{\rm min}\equiv m_{\rm thermal}^{\rm min}$ is the lower bound on the warm dark matter mass. Furthermore, $\omega_{\rm dm}$ is the abundance of self-interacting dark matter. If it provides only a fraction of the observed dark matter abundance, one should use the corresponding lower limit on the warm dark matter mass for mixed cold/warm dark matter \cite{Baur:2017stq, Boyarsky:2008xj}, provided the condition $z_f\gtrsim 10^6$ holds (see \cite{Buen-Abad:2018mas} for a discussion of sub-dominant cannibal dark matter). 

\begin{figure}[t]
\begin{center}
\includegraphics[width=0.49\textwidth]{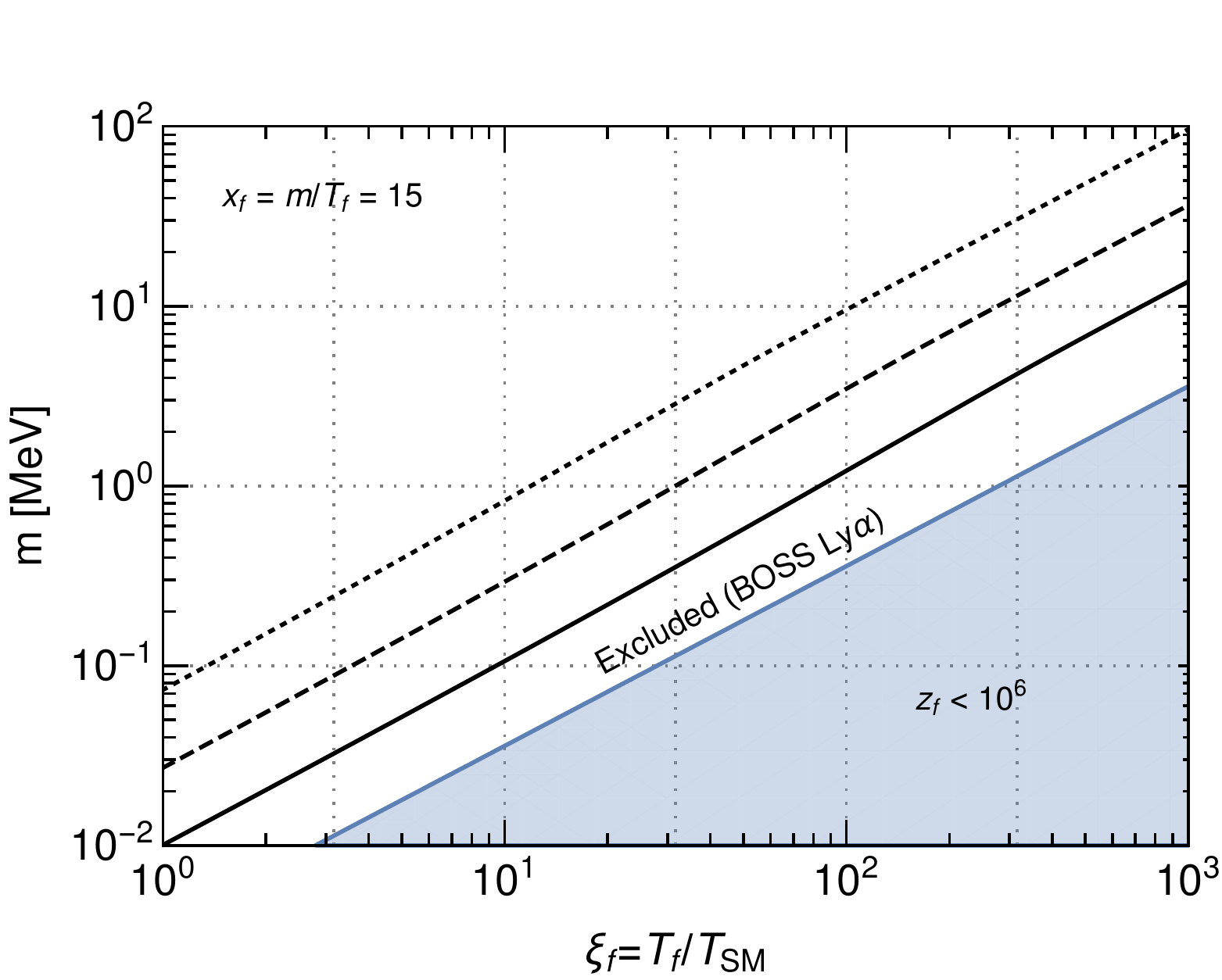}
\end{center}
\caption{\label{fig:sidm_massbound}%
\small Lower bound on the dark matter mass for strongly self-interacting dark matter from Lyman-$\alpha$ constraints corresponding to $m_{\rm WDM}^{\rm min}=1,2,4$\,keV (solid, dashed, dotted lines). Dependence on $\xi_f=T_f/T_{SM}$ for $x_f=m/T_f=15$. 
Within the blue region the modes relevant for Lyman-$\alpha$ observations would enter the horizon before number-changing interactions have frozen out.
}
\end{figure}

For the case in which all of dark matter is self-interacting we illustrate the lower bound on the dark matter mass in Fig.\,\ref{fig:sidm_massbound}, assuming $x_f=15$ and using the number of degrees of freedom for the entropy $g_{*S}$ adapted from ~\cite{Laine:2015kra}.
It ranges from $m\geq 0.29(0.83)$\,MeV for $\xi_f=10$ to
$m\geq 36(96)\,$MeV for $\xi_f=10^3$, when using $m_{\rm WDM}^{\min}=2(4)$\,keV. Note that the assumption $z_f>10^6$ is satisfied, i.e. freeze-out takes place \emph{before} the relevant modes for Lyman-$\alpha$ observations have entered the horizon. The region in which this is not the case is shown by the blue shaded area in Fig.\,\ref{fig:sidm_massbound}.
\begin{figure}[t]
\begin{center}
\includegraphics[width=0.49\textwidth]{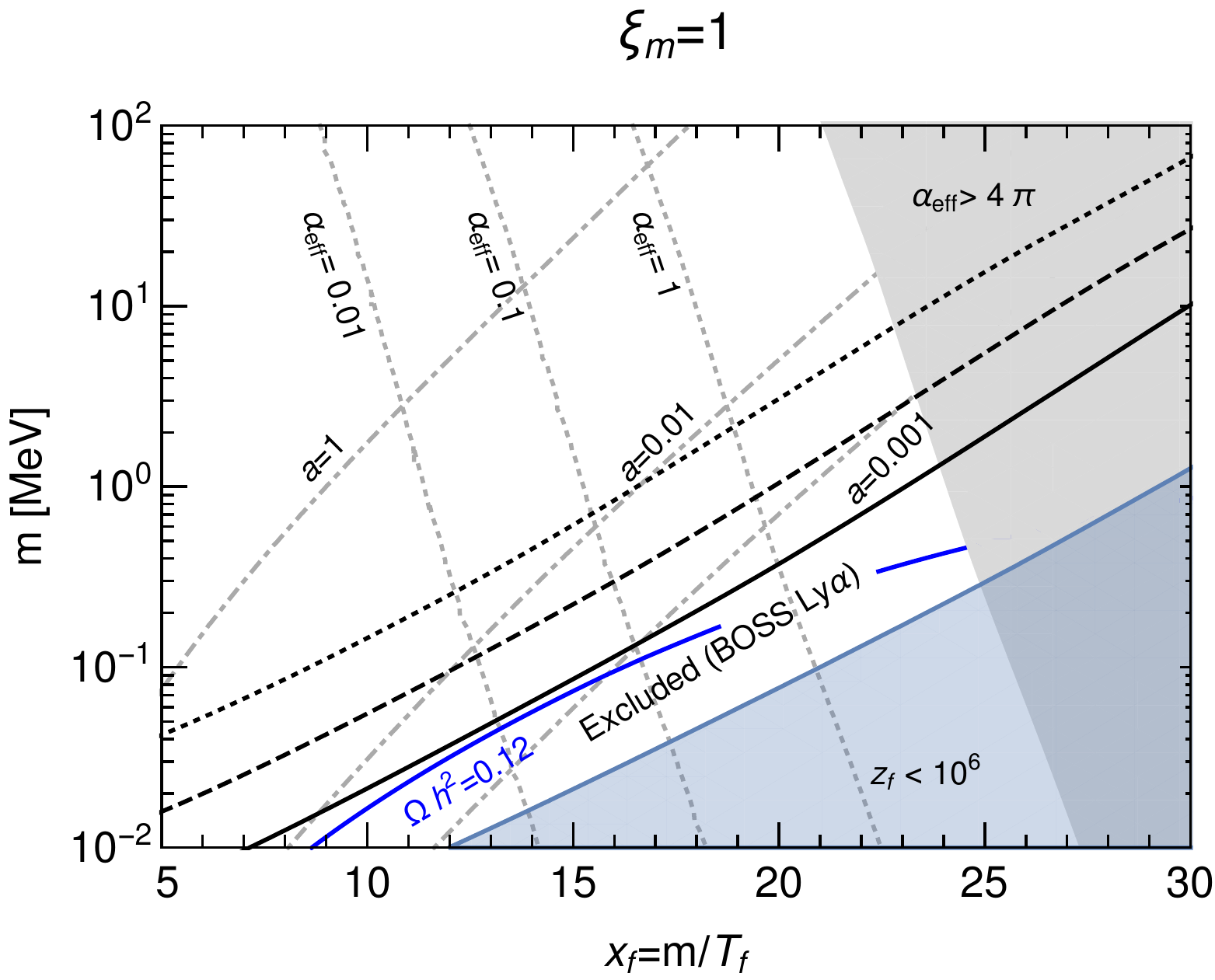}
\includegraphics[width=0.49\textwidth]{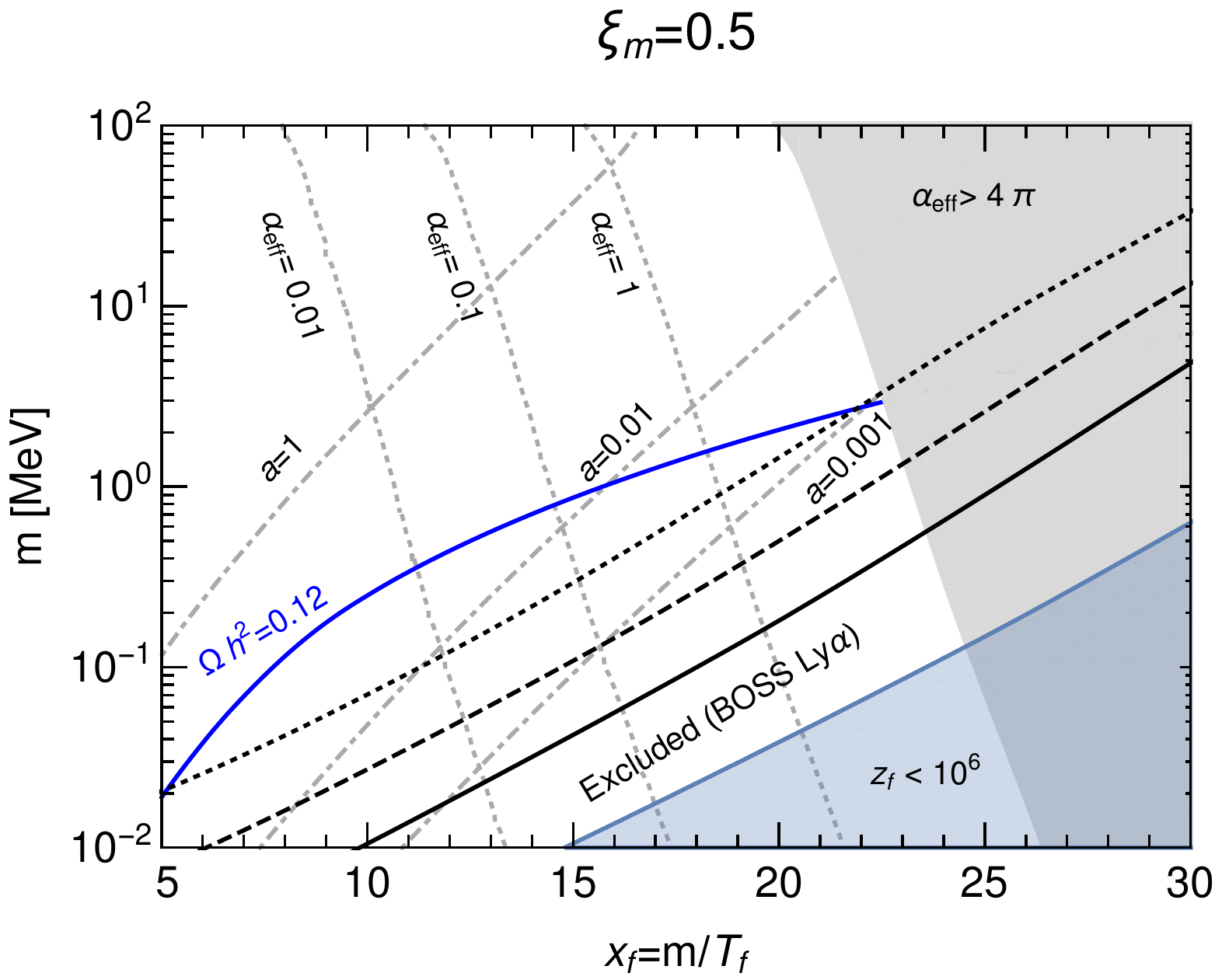}
\end{center}
\caption{\label{fig:sidm_massbound2}%
\small Lower bound on the dark matter mass for strongly self-interacting dark matter from Lyman-$\alpha$ constraints corresponding to $m_{\rm WDM}^{\rm min}=1,2,4$\,keV (black solid, dashed, dotted lines). The blue line indicates the parameters that yield the observed dark matter abundance for glueball dark matter~\cite{Soni:2016gzf}, and the gray dotted lines indicate the corresponding effective coupling strength $\alpha_{\rm eff}$. Grey dot-dashed lines correspond to $\sigma_{2\to2}/m=1$barn/GeV for $a=1,0.01,10^{-3}$ (see text). The blue shaded region is as in Fig.\,\ref{fig:sidm_massbound2}.
}
\end{figure}
If we assume that $T(a)\propto 1/\ln(a)$ for $1<x=m/T<x_f$, and $T/T_{\rm SM}\equiv \xi_m$ when $T=m$,
the temperature ratio at freeze-out is fixed, and given by~\cite{Carlson:1992fn,deLaix:1995vi,Soni:2016gzf}
\begin{equation}
  \xi_f = \frac{T_f}{T_{\rm SM}} \approx \xi_m \frac{e^{x_f/3}}{x_f}\,\left(\frac{g_{*S}(m/(\xi_fx_f))}{g_{*S}(m/\xi_m)}\right)^{\frac13}\,.
\end{equation}
The resulting lower bound on $m$ is shown in Fig.\,\ref{fig:sidm_massbound2}  for the case in which dark matter is thermalized with the SM bath for $T>m$, i.e. $\xi_m=1$ (left panel) and for $\xi_m=0.5$ (right panel). We also indicate the value for which the relic density would match the observed value within the scenario of glueball dark matter~\cite{Soni:2016gzf},
and the corresponding dimensionless effective coupling strength $\alpha_{\rm eff}\equiv (\langle\sigma v^2\rangle_{3\to 2}m^5)^\frac13$~\cite{Hochberg:2014dra}. In addition, we show contours for which the $2\to2$ scattering cross section divided by the mass
\begin{equation}
\frac{\sigma_{2\to 2}}{m}=\frac{a^2\alpha_{\rm eff}^2}{m^3}
\end{equation}
is equal to $1$\,barn/GeV when assuming $a=1,0.01,10^{-3}$, respectively~\cite{Hochberg:2014dra}.

\subsection{Dark matter interacting with dark radiation}

We consider a dark sector containing either one or two species of dark matter particles,
and a dark radiation component. Before briefly reviewing microphysical realizations, we describe
the impact on cosmological perturbations. Let us first consider the case with a single dark matter
component. An interaction between dark matter and dark radiation
leads to a drag force $\frac{d}{d\eta}\vec v=-a\Gamma\vec v$ on (non-relativistic) interacting dark matter particles moving through the
bath of dark radiation, described by a friction coefficient $\Gamma$. Here $d\eta=dt/a$ is related to the conformal time.
The impact of this drag force on cosmological perturbations can be captured by the evolution equations \cite{Buen-Abad:2015ova,Lesgourgues:2015wza,Buen-Abad:2017gxg}
\bea\label{eq:pert}
  \dot\delta_{\rm idm} &=& -\theta_{\rm idm} +3\dot\phi\,, \nn\\
  \dot\theta_{\rm idm} &=& -{\cal H}\theta_{\rm idm} +a\Gamma(\theta_{\rm dr}-\theta_{\rm idm})+k^2\psi\,, \\
  \dot\delta_{\rm dr} &=& -\theta_{\rm dr} +4\dot\phi\,, \nn\\
  \dot\theta_{\rm dr} &=& \frac{k^2}{4}\delta_{\rm dr} +\frac34\frac{\rho_{\rm idm}}{\rho_{\rm dr}} a\Gamma(\theta_{\rm idm}-\theta_{\rm dr})+k^2\psi \,,
\eea
where 
$\rho_i,\delta_i,\theta_i$ are the homogeneous
background density, the density contrast, and the velocity divergence of species $i$. Furthermore, we assume that self-interactions of the dark
radiation component erase higher moments of its distribution function, such that it behaves as an ideal fluid completely described by $\delta_{\rm dr}$
and $\theta_{\rm dr}$. Apart from the usual $\Lambda$CDM parameters and the friction term $\Gamma$, the model is described by the density of dark radiation, which can be
captured by the parameter $\Delta N_{fluid}=\rho_{\rm dr}/\rho_\nu$, where $\rho_\nu\equiv \frac78\frac{\pi^2}{15}(\frac{4}{11})^{4/3}T_\gamma^4(1+z)^4$ and
$T_\gamma$ is the average CMB temperature.

In addition, one may consider the possibility that only a fraction $f_{idm}\in[0,1]$ of the total dark matter abundance is interacting with dark radiation \cite{Chacko:2016kgg,Buen-Abad:2017gxg}.
In this case we need to complement the evolution equations by a standard CDM component that interacts only gravitationally,
and replace $\delta_{\rm cdm}\to f_{idm}\delta_{\rm idm}+(1-f_{idm})\delta_{\rm cdm}$ on the right-hand side of the evolutions equations for
the metric perturbations~\cite{Ma:1995ey}.

The setup described above captures a large variety of microscopic realizations of interacting dark matter models, see e.g. \cite{Aarssen:2012fx,Blennow:2012de,Diamanti:2012tg,Chacko:2015noa,Buen-Abad:2015ova,Lesgourgues:2015wza,Cyr-Racine:2015ihg}.
Let us briefly review one of them, following \cite{Buen-Abad:2015ova,Lesgourgues:2015wza}. 
In this setup dark radiation is composed of non-abelian gauge bosons $\gamma_d$ of a dark gauge group
$SU(N)_d$ with gauge coupling constant $g_d\ll 1$. Interacting dark matter is a stable particle $\chi$ charged 
under the dark gauge group. The non-Abelian symmetry leads to self-interactions of $\gamma_d$, which ensure the ideal fluid behavior, with negligible viscosity. 
If the dark radiation was in equilibrium with the thermal bath of Standard Model particles until electroweak scales, its abundance is given by 
$\Delta N_{fluid}\simeq 0.07(N^2-1)$ \cite{Buen-Abad:2015ova}. The drag force arises from Compton scattering $\chi\gamma_d\to\chi\gamma_d$ in the
dark sector. For a Dirac fermion $\chi$ in the fundamental representation, the $t$-channel exchange of a dark gauge boson yields the 
dominant contribution to the drag force \cite{Lesgourgues:2015wza}
\be
  \Gamma = \frac{\pi}{9}(N^2-1)\alpha_d^2\ln\alpha_d^{-1}\,\frac{T_d^2}{m_\chi}\;,
\ee
where $\alpha_d=g_d^2/(4\pi)$ and $m_\chi$ is the dark matter mass. The temperature-dependence $\propto T_d^2$ of the interaction rate can be parameterized by $\Gamma=\Gamma_0/a^2$ with $\Gamma_0$
given by the rate today. 
This scaling relies on the dependence of the scattering amplitude ${\cal M}\propto 1/q^2$ on the momentum transfer $q$.

\begin{figure}[t]
\begin{center}
\includegraphics[width=0.49\textwidth]{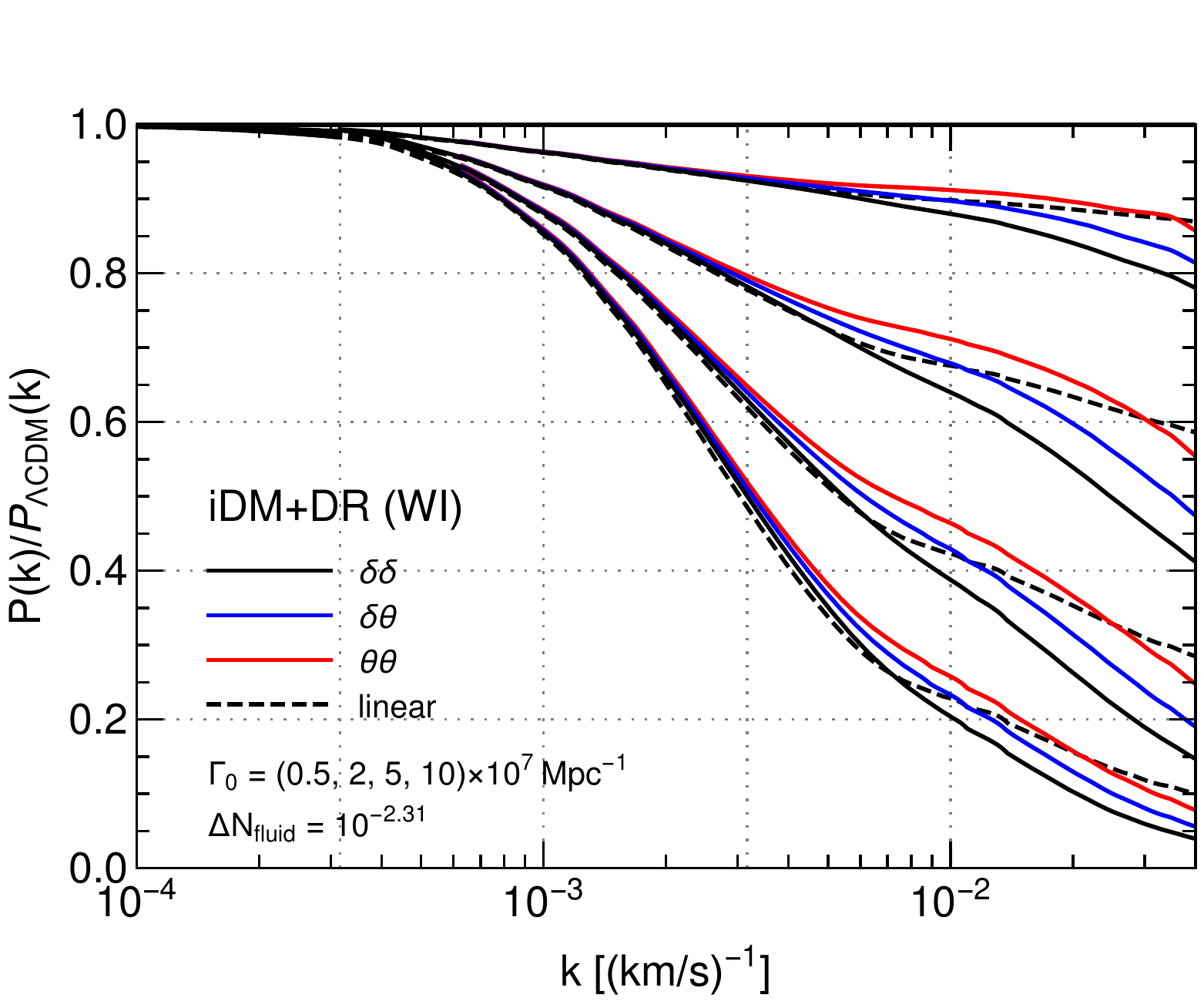}
\includegraphics[width=0.49\textwidth]{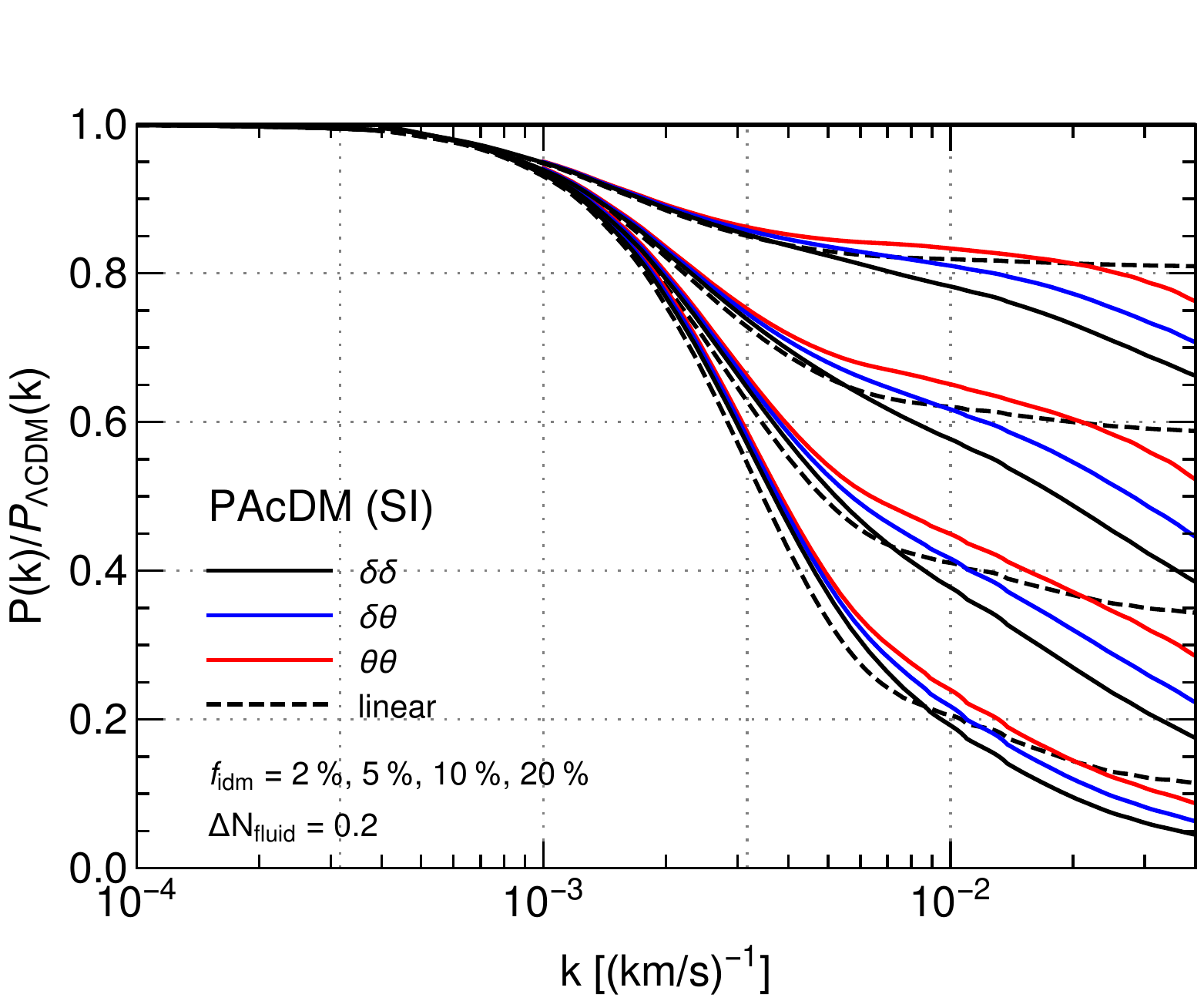}
\end{center}
\caption{\label{fig:powerspec_SIWI}%
\small Power spectra for dark matter interacting with dark radiation, normalized to the $\Lambda$CDM power spectrum. Left panel: weakly interacting (WI) limit with $f_{idm}=100\%$ and for $\Gamma_0\times 10^7$\,Mpc\,$=0.5,2,5,10$ (from top to bottom). Right panel: strongly interacting limit (SI) corresponding to the PAcDM model with $\Gamma\gg H$ and a fraction of interacting dark matter $f_{idm}=2\%, 5\%, 10\%, 20\%$ (from top to bottom). Black dashed lines show the ratio of linear power spectra, and solid lines correspond to non-linear spectra at $z=3.6$ for the density, the velocity divergence, and the cross spectrum, respectively. All other parameters are fixed to those of the SI log and WI log benchmark models (see Tab.\,\ref{tab:benchmark_param}).  To obtain the initial power spectra for the interacting dark matter models, we modified the CLASS code~\cite{2011arXiv1104.2932L} (see App.~\ref{sec:class} for details).
}
\end{figure}

In \cite{Lesgourgues:2015wza, Buen-Abad:2017gxg} it was argued that this model can alleviate the
tension between constraints on the Hubble constant $H_0$ from direct measurements \cite{Riess:2016jrr} and Planck constraints \cite{Ade:2015xua}, as well as the tension in $\sigma_8$ considering lensing and clustering observations \cite{Heymans:2013fya,Ade:2015fva,Joudaki:2017zdt,Abbott:2017wau}, provided the drag force is of order $1/\Gamma_0 \sim {\cal O}(10^7)$\,Mpc, corresponding to $\alpha_d\sim 10^{-8}$ for a TeV-scale
mass $m_\chi$. These relatively weak interactions freeze out when the perturbation modes observed in large-scale structure enter the horizon, and lead to a
suppression of the power spectrum on smaller scales. The linear and non-linear (2-loop) power spectra for $z=3.6$ are shown in Fig.\,\ref{fig:powerspec_SIWI} (left panel) for four values of $\Gamma_0$ in the range $5\cdot 10^{-8}$Mpc$^{-1}$ to $10^{-6}$Mpc$^{-1}$.

In \cite{Chacko:2016kgg} a different setup was considered, termed partially acoustic dark matter (PAcDM), which can be mapped to the same set of evolution equations
as discussed above, for $f_{idm}<1$  \cite{Buen-Abad:2017gxg}. In this case the rate $\Gamma$ can be much larger,
such that the interacting component of dark matter is tightly coupled to the dark radiation (which occurs for $\Gamma\gg H$). In this case Compton scatterings
equalize the velocities of the idm and dr components, similar to the baryon-photon plasma, and the rate $\Gamma$ drops out of the evolution equations. Therefore, this strongly-interacting limit is captured also by only two additional free parameters, $f_{idm}$ and $\Delta N_{fluid}$.
 Similar to the weakly interacting case, the interactions lead to a scale-dependent suppression of the power spectrum. An example for various values in the range $f_{idm}=2\%-10\%$ is shown in Fig.\,\ref{fig:powerspec_SIWI} (right panel).
In \cite{Buen-Abad:2017gxg} it has been argued that the tension between measurements of $H_0$ and $\sigma_8$ can be alleviated as compared to the best-fit $\Lambda$CDM
model for values of $f_{\rm idm}$ in the few-percent range. It seems difficult to relieve both the $H_0$ and the $\sigma_8$ tension at the same time within this setup~\cite{Raveri:2017jto}.

The scale-dependent suppression of the power spectrum is potentially testable with Lyman-$\alpha$ data. 
In \cite{Krall:2017xcw} SDSS Lyman-$\alpha$ data~\cite{McDonald:2004eu} from 2004 were applied for the weakly interacting limit, finding that
the preference over $\Lambda$CDM is somewhat reduced. On the other hand, Ref.\,\cite{Pan:2018zha} argues that adding Lyman-$\alpha$ data may favor interacting
dark matter, taking as an indicator the
constraints on the slope and amplitude of the linear power spectrum that were derived under the assumption of $\Lambda$CDM in \cite{Palanque-Delabrouille:2015pga}.

\begin{table}
\begin{center}
\begin{tabular}{ l|l|l|l|l|l|l|l|l|l||l|l| }
\cline{2-12}
& $100\omega_b$ & $n_s$ & $\tau_{reio}$ & $H_0$ & $\ln 10^{10}A_s$ & $\omega_{dm}^{\rm tot}$ & $\Delta N_{fluid}$ & $10^7\Gamma_0$ & $f_{\rm idm}$ & $\sigma_8$ & $\Delta\chi^2_{\rm eff}$\\
\hline
\hline
\multicolumn{1}{|c|}{CDM} & 2.245 & 0.9656 & 0.04887 & 68.87 & 3.023 & 0.1168 & 0 & 0 & 0 & 0.7933 & 0\\ \hline
\multicolumn{1}{|c|}{SI lin} & 2.242 & 0.9701 & 0.06118 & 69.13 & 3.056 & 0.1235 & $0.2^*$ & $\gg H_0$ & 0.0139 & 0.7734 & -10.3\\ \hline
\multicolumn{1}{|c|}{SI log} & 2.231 & 0.9628 & 0.05827 & 67.98 & 3.047 & 0.1185 & $10^{-2.81}$ & $\gg H_0$ & 0.0479 & 0.7588 & -20.0\\ \hline
\multicolumn{1}{|c|}{WI lin} & 2.249 & 0.9708 & 0.05915 & 70.01 & 3.050 & 0.126 & 0.369 & 1.097 & 1 & 0.7721 & -14.3\\ \hline
\multicolumn{1}{|c|}{WI log} & 2.228 & 0.9625 & 0.05815 & 67.84 & 3.047 & 0.119 & $10^{-2.31}$ & 2.272 & 1 & 0.7565 & -22.2\\ \hline
\multicolumn{1}{|c|}{PKK} & 2.266 & 0.9833 & 0.09749 & 72.77 & 3.104 & 0.1345 & 0.8629 & 1.353 & 1 & 0.7969 & $<0$ \\ \hline
\end{tabular}
\caption{\label{tab:benchmark_param}%
\small 
Parameters describing interacting dark matter benchmark models, see \cite{Buen-Abad:2017gxg} (and \cite{Pan:2018zha} for the last line) for details.
In the last line we show the values of $\Delta\chi^2_{\rm eff}$ relative to CDM taking into account
Planck 2015 TT/TE/EE data combined with BAO, $H_0$, and lensing/clustering data as described in \cite{Buen-Abad:2017gxg}. 
In the SI (lin) model we fixed $\Delta N_{fluid}=0.2$, because the likelihood analysis of \cite{Buen-Abad:2017gxg} found only an upper limit
$\Delta N_{fluid}<0.506\, (95\%{\rm C.L.})$. $\Gamma_0$ is given Mpc$^{-1}$ and $H_0$ in km/s/Mpc.
}
\end{center}
\end{table}
\begin{table}
\begin{center}
  \begin{tabular}{ | c | c | c | c | c | c | c | c | c | }
    \hline
	& prior & CDM & SI lin & SI log & WI lin & WI log & PKK & mock \\
    \hline
    \hline
    $A$ & $>0$ 	
& $0.428$ 	& $0.443$ 	& $0.521$ 	& $0.437$  	& $0.482$ & $0.417$ & $0.428$\\ 
    \hline
    $\beta_F$	&
& $2.37$  	& $2.37$ 	& $2.91$  	& $2.43$ 	& $2.69$ & $2.41$ & $2.37$\\ 
    \hline
    $\alpha_{\rm bias}$ 	     &  	
& $1.40$ 	& $1.41$ 	& $0.11$ 	& $1.37$  	& $0.76$& $1.38$ & $1.11$ \\ 
    \hline
    $\beta_{\rm bias}$ 	     &  	
& $0.43$ 	& $0.44$ 	& $6.36$ 	& $0.66$  	& $1.80$ & $0.61$ & $-0.37$\\ 
    \hline
    $\alpha_{\rm c.t.}$  $[(km/s)^{-1}]$   	& $<0$ 		
& $-5042$   	& $-4602$ 	& $-848$ 	  & $-3486$ & $-1094$  & $-4185$ & $-4287$\\ 
    \hline
    $\beta_{\rm c.t.}$ 	&
& $4.34$ 	 & $4.29$ 	& $5.63$ 	 & $4.29$ 	& $4.90$ & $4.34$ & $4.88$ \\ 
    \hline
    \hline
    $\chi^2$ 	&
& $231.7 / 210$ 	& $233.9 / 210$ 	& $245.2 / 210$	 & $225.1 / 210$ & $231.8 / 210$  & $225.9/210$  & $125.9/204$ \\
    \hline
  \end{tabular}
\caption{\label{tab:fit}%
All parameters that are varied in the fit and their corresponding best fit points. The fits uses six parameters and the data from the six redshift bins $[2.6,3.6]$ and the data BOSS1.
The last column shows the best fit values obtained from the simulated mock data for a $\Lambda$CDM model (see Sec.\,\ref{sec:num}).
\small 
}
\end{center}
\end{table}

In order to test the compatibility of interacting dark matter with the Lyman-$\alpha$ data from BOSS~\cite{Palanque-Delabrouille:2013gaa}, we consider the best-fit
benchmark models provided in \cite{Buen-Abad:2017gxg} for both the weakly-interacting (WI) and strongly-interacting (SI) limit, and considering a linear prior $\Delta N_{fluid}>0.07$ (lin) or a flat logarithmic prior on $\Delta N_{fluid}$ allowing for smaller dark radiation densities (log). In addition, we consider a benchmark model
from the analysis of Ref.\,\cite{Pan:2018zha}, for which the slope and amplitude of the linear power spectrum indicate a good agreement with Lyman-$\alpha$ data (denoted by ``PKK''). The input parameters and values of $\Delta\chi^2$ obtained from CMB, BAO, $H_0$ and lensing/clustering data relative to the best-fit $\Lambda$CDM model are given in Tab.\,\ref{tab:benchmark_param}. We note that in~\cite{Pan:2018zha} it has been argued that $\Delta\chi^2$ is reduced when allowing for more freedom in the SZ cluster mass function.

\begin{figure}[t]
\centering
\begin{minipage}{\textwidth}
\centering
\includegraphics[width=0.48\textwidth]{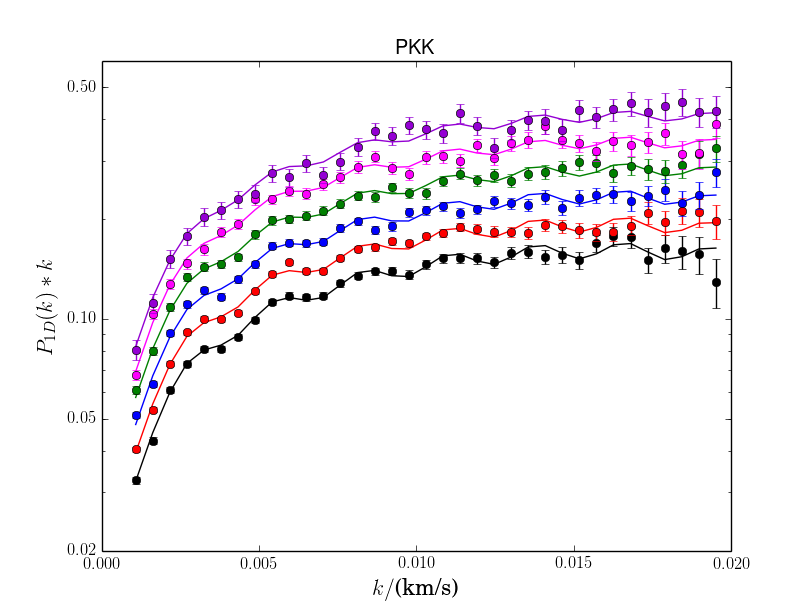}
\includegraphics[width=0.48\textwidth]{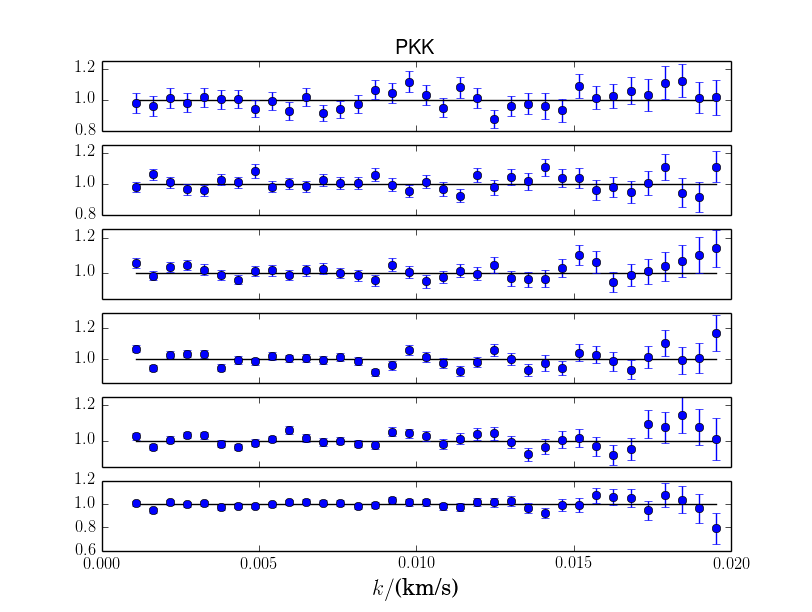}
\caption{\small Best fit of the Lyman-$\alpha$ flux power spectrum for the `PKK' interacting dark matter benchmark model~\cite{Pan:2018zha} to the BOSS1 data. On the left, we show our theoretical model prediction for the Lyman-$\alpha$ flux power spectrum $P_{1D}(k,z)$ (solid lines) as a function of the wavevector along the line-of-sight, $k$, and for redshifts $z=[2.6,2.8,3.0,3.2,3.4,3.6]$ (from bottom to top) in comparison to the BOSS1 data (dots). The corresponding residuals of the model fit for the same redshifts  are shown on the right.}
\label{fig:KaplinghatBOSS}
\end{minipage}
\begin{minipage}{\textwidth}
\centering
\includegraphics[width=0.48\textwidth]{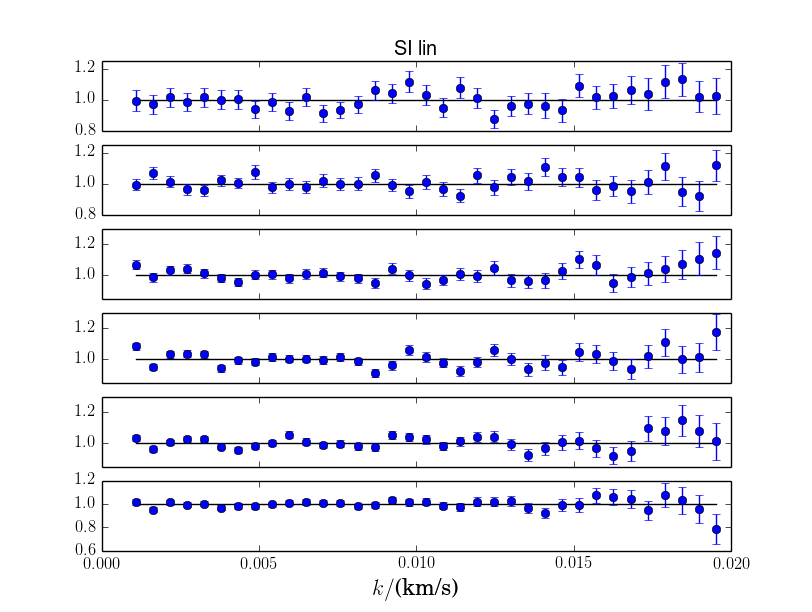}
\includegraphics[width=0.48\textwidth]{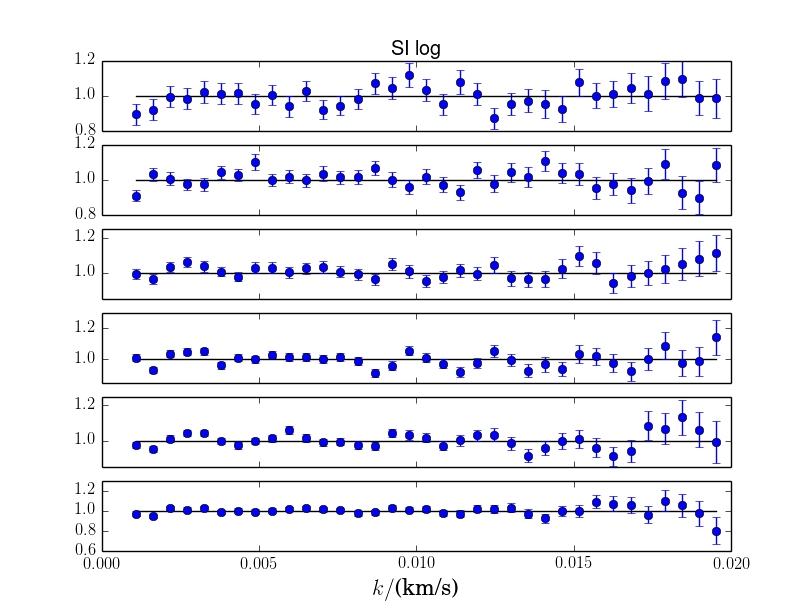}
\includegraphics[width=0.48\textwidth]{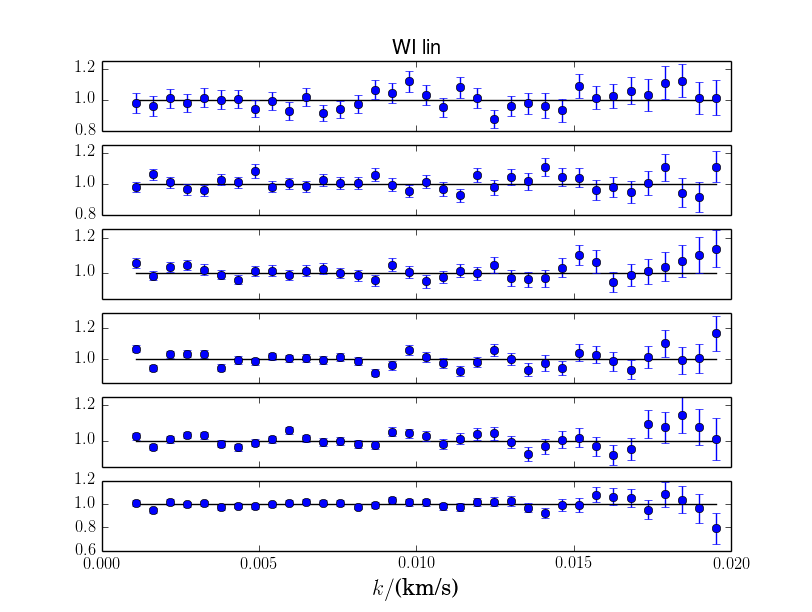}
\includegraphics[width=0.48\textwidth]{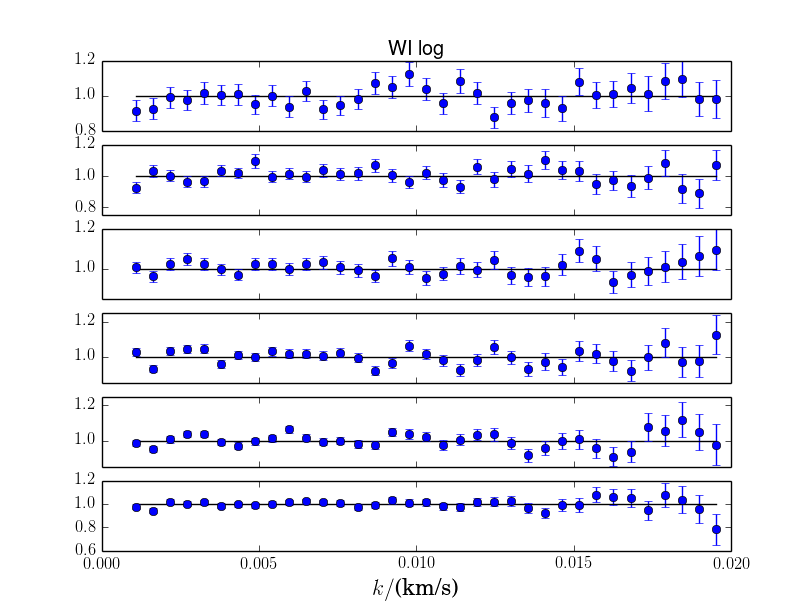}
\caption{\small BOSS Lyman-$\alpha$ flux power spectrum over the model predictions for the benchmarks SI log/lin and WI log/lin. The six subplots show the redshifts $z=[2.6, 2.8,3.0,3.2,3.4,3.6]$ from bottom to top.}
\label{fig:residuals}
\end{minipage}
\end{figure}

We show our results for the best-fit values of the various model parameters based on BOSS Lyman-$\alpha$ data~\cite{Palanque-Delabrouille:2013gaa}, as well as the resulting $\chi^2$,  in Tab.\,\ref{tab:fit},  for the case with a redshift dependence in the bias function $\beta$, and the BOSS1 data set.
For comparison, we also include the best-fit $\Lambda$CDM model taken from \cite{Buen-Abad:2017gxg}.
As mentioned before, some of the parameters in our model are only marginally relevant for the fit and are fixed to the values as given in Tab.~\ref{tab:fixed}. The resulting flux power spectrum is shown in Fig.\,\ref{fig:KaplinghatBOSS} (left panel) for the ``PKK''
benchmark model. The residuals  with respect to BOSS1 data are shown in Fig.\,\ref{fig:KaplinghatBOSS} and Fig.\,\ref{fig:residuals} for the interacting dark matter models.
In addition, the best-fit $\chi^2$ values for BOSS2 data as well as for different assumptions on the bias $\beta$ are collected in Tab.\,\ref{tab:final1}.
\begin{table}
\begin{center}

\begin{tabular}{ c|c|c|c|c|c|c| }
\cline{2-7}
& \multicolumn{3}{c|}{BOSS1} & \multicolumn{3}{c|}{BOSS2} \\
\cline{2-7}
& $\beta_{\rm bias} = 0$ & $\beta_{\rm bias}$ arbitrary & $\beta_{\rm bias} < 1$ 
& $\beta_{\rm bias} = 0$ & $\beta_{\rm bias}$ arbitrary & $\beta_{\rm bias} < 1$ \\
\hline
\hline
\multicolumn{1}{|c|}{CDM} 
& 239.7 & 231.7 & (231.7) 
& 191.8 & 170.8 & (170.8)\\ \hline
\multicolumn{1}{|c|}{SI lin} 
& 241.7 & 233.9 & (233.9) 
& 192.5 & 171.3 & (171.3)\\  \hline
\multicolumn{1}{|c|}{SI log} 
& 262.1 & 245.2 & 254.8
& 210.8 & 182.9 & 196.8\\ \hline
\multicolumn{1}{|c|}{WI lin} 
& 237.5 & 225.1 & (225.1) 
& 194.7 & 167.7 & (167.7) \\ \hline
\multicolumn{1}{|c|}{WI log} 
& 250.6 & 231.8 & 234.4
& 206.3 & 173.9 & 182.6 \\ \hline
\multicolumn{1}{|c|}{PKK} 
& 238.4 & 225.9 & (225.9) 
& 195.4 & 168.5 & (168.5)\\ \hline
\end{tabular}
\caption{\label{tab:final1}%
\small 
Best-fit value of $\chi^2$ for the benchmark models from Tab.\,\ref{tab:benchmark_param} for BOSS1 and BOSS2 Lyman-$\alpha$ data, and for various assumptions on the bias parameter $\beta_{\rm bias}$, respectively.
 For the number in brackets, the prior $\beta_{\rm bias}<1$ is ineffective since the best fit point is consistent with the prior. The fit uses the six redshift bins $[2.6,3.6]$ amounting to $210$ degrees of freedom. 
}
\end{center}
\end{table}

We find that the interacting models `WI log' and `SI log' indicating the strongest preference over $\Lambda$CDM from LSS and CMB data are slightly disfavored by BOSS Lyman-$\alpha$ observations, when assuming a constant bias $\beta$. Allowing for a redshift dependence relaxes the tension for the `WI log' model. On the other hand, the `WI lin' and `SI lin' benchmark models are compatible with Lyman-$\alpha$ observations, or even slightly favored, depending on the data set and assumptions on the bias.
Finally, we find that the `PKK' benchmark model is favored by BOSS Lyman-$\alpha$ data, in line with the indications discussed in \cite{Pan:2018zha}, although the improvement compared to $\Lambda$CDM seems to be moderate.

\begin{figure}[t]
\begin{center}
\includegraphics[width=0.49\textwidth]{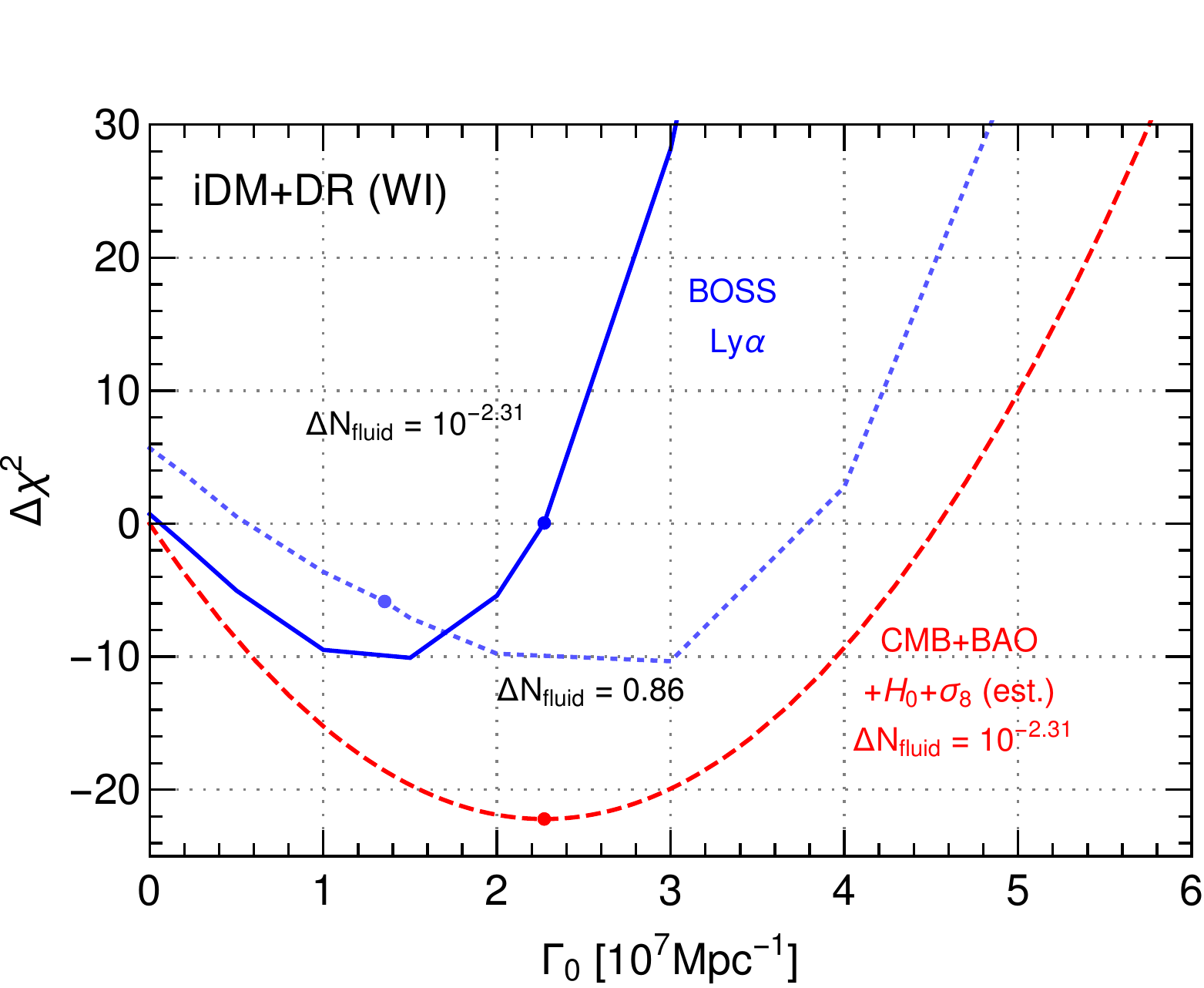}
\includegraphics[width=0.49\textwidth]{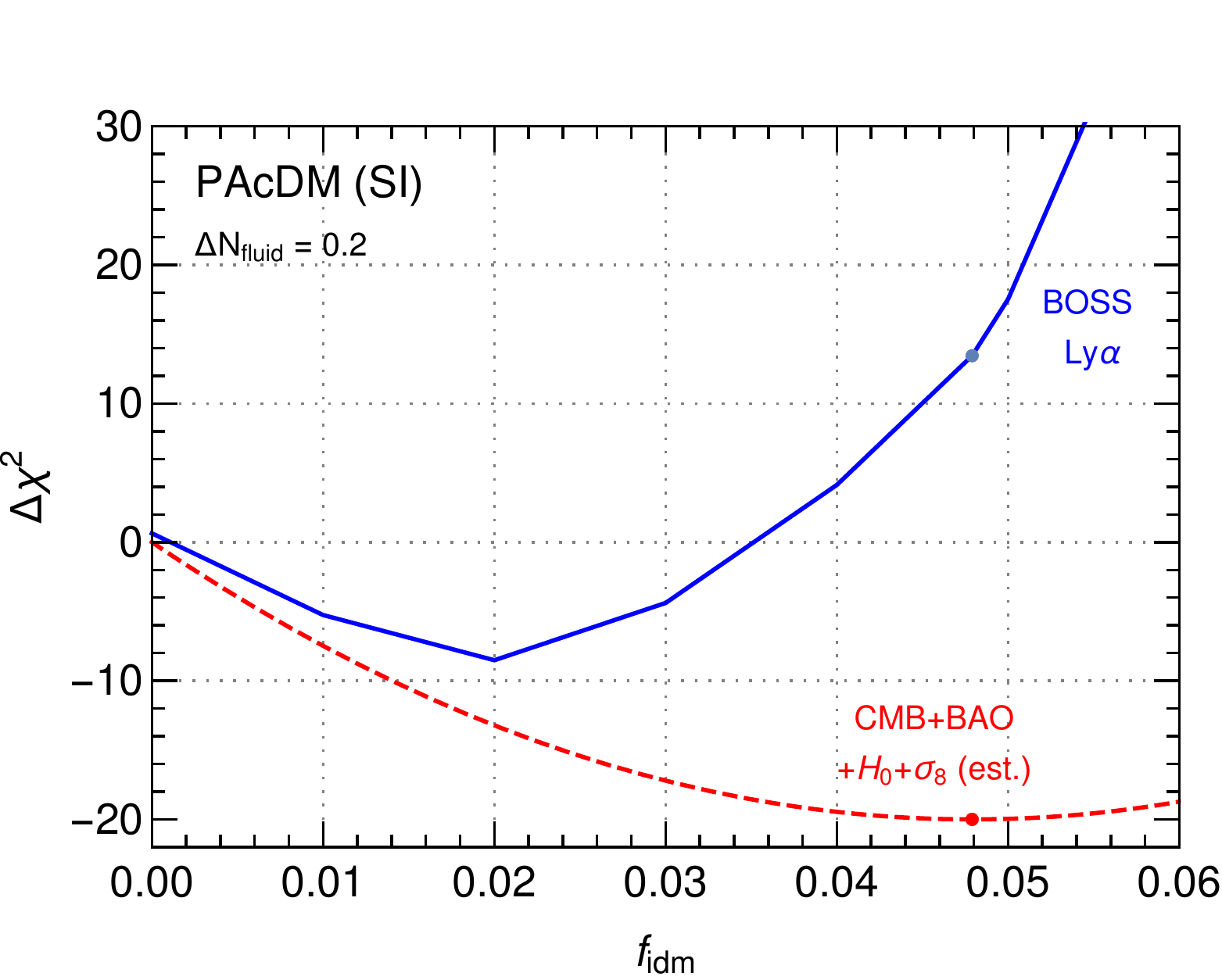}
\end{center}
\caption{\label{fig:slices}%
\small 
Best-fit value of $\Delta\chi^2$ (blue lines) for BOSS1 data and redshift-dependent bias $\beta$, relative to the $\Lambda$CDM model  from Tab.\,\ref{tab:benchmark_param}, for one-dimensional slices of the parameter space in the WI limit (left panel) and SI limit (right panel). The red dashed line corresponds to a Gaussian estimate of $\Delta\chi^2$ from the combination of CMB and LSS data considered in \cite{Buen-Abad:2017gxg}.
In the WI limit we vary the interaction rate $\Gamma_0$ and keep all other cosmological parameters fixed to the values as for the WI log model (blue solid line) or the PKK model (blue dotted line), respectively. In the SI limit we vary $f_{idm}$ with all other parameters being identical to the SI log model, see Tab.\,\ref{tab:benchmark_param}. The dots indicate the values for the corresponding benchmark models.
}
\end{figure}

In addition, we consider one-dimensional slices of the full parameter space by varying the dark matter interaction rate $\Gamma_0$ for the WI limit, or the interacting dark matter fraction $f_{idm}$ for the SI limit, respectively, while keeping all other cosmological parameters fixed. The resulting $\Delta\chi^2$, relative to the $\Lambda$CDM benchmark from Tab.\,\ref{tab:benchmark_param}, are shown in Fig.\,\ref{fig:slices} for the BOSS1 Lyman-$\alpha$ data set and when assuming redshift-dependent bias. Interestingly, the Lyman-$\alpha$ data show a slight preference for non-zero interaction rate in the WI case, and non-zero fraction of interacting dark matter in the SI case, albeit for slightly smaller values as compared to CMB and LSS data. Nevertheless, these findings indicate that models of interacting dark matter addressing the $H_0$ and $\sigma_8$ tension can be compatible with Lyman-$\alpha$ data.

We note that the present conservative analysis should however be taken as indicative. The  method of deriving  Lyman-$\alpha$ constraints discussed here allows in principle for a joint likelihood analysis of the full parameter space based on Monte Carlo sampling techniques, which is left to future work.

\section{Conclusion \label{sec:dis}} 

Observations of the Lyman-$\alpha$ forest provide a powerful probe of models with a power spectrum that is suppressed on small scales as compared to $\Lambda$CDM. Extracting constraints from measurements of the Lyman-$\alpha$ flux power spectrum requires to model the dynamics of the intergalactic medium, as well as the underlying density field. In this work we have considered an effective model for the flux power spectrum on large, quasi-linear scales $k<0.02($km/s$)^{-1}$ relevant for BOSS Lyman-$\alpha$ data~\cite{Palanque-Delabrouille:2013gaa}, intended to yield an approximate, conservative estimator of Lyman-$\alpha$ constraints at relatively low computational cost. The impact of the intergalactic medium on large scales, as well as the sensitivity to smaller, highly non-linear scales due to the line-of-sight integration, are encapsulated into a number of nuisance parameters that are determined by fitting to the data. The exquisite quality of BOSS data allows us to extract conservative constraints, even when allowing for significant freedom in the theoretical model.
We employed a simplified Gaussian treatment of observational uncertainties based on the published results, and adding the quoted statistical and systematic uncertainties in quadrature, as a first step here. Ultimately, the theoretical model should be applied to the full likelihood function used by the BOSS collaboration.

We validated the fitting model by comparing to simulated mock data, finding excellent agreement on scales measured by BOSS.
Building upon this, we applied our fitting procedure to a number of models for self-interacting dark matter that have been considered in the literature. For models of strongly self-interacting dark matter, that feature number-changing self-annihilations, we derive a lower bound on the dark matter mass depending on the dark matter temperature $T_f$ when number-changing interactions freeze out. For example, we find a conservative lower limit $m\gtrsim $\,MeV for $T_f=m/15=100T_{\rm SM}$.

In addition, we considered models with a dark radiation component, that interacts with dark matter, including the so-called partially acoustic dark matter scenario (PAcDM) as well as a related proposal with weaker interaction strength, but only one dark matter component. These types of setups have been argued to address the $H_0$ and $\sigma_8$ tensions within $\Lambda$CDM. Applying our conservative analysis we find that they are compatible with BOSS Lyman-$\alpha$ data, although the preference over $\Lambda$CDM inferred from CMB and LSS data according to the analysis of~\cite{Buen-Abad:2017gxg} is slightly reduced. Nevertheless, our findings indicate that a combined fit could lead to an better overall agreement for somewhat smaller interaction strength and/or fraction of interacting dark matter, with a preference over $\Lambda$CDM persisting in this case.


\section*{Acknowledgements}

We thank M.~Kaplinghat and Z.~Pan for helpful discussions.
LS and ST are supported by the Natural Sciences and Engineering Research Council of Canada. TK acknowledges support by the German Science Foundation (DFG) within the Collaborative Research Center (SFB) 676 `Particles, Strings and the Early Universe'.

\appendix

\section{Non-linear power spectrum and CLASS implementation \label{sec:class}} 

In order to obtain the linear matter power spectrum for interacting dark matter coupled to dark radiation, we implement
the perturbation equations \eqref{eq:pert} in CLASS \cite{2011arXiv1104.2932L}. We cross-checked the implementation of the
dark radiation component which behaves as an ideal fluid based on \cite{Chacko:2015noa,Baumann:2015rya}, and of the interacting
dark matter component by verifying that the linear matter power spectra agree with \cite{Buen-Abad:2017gxg}.

The non-linear matter power spectrum was obtained using the implementation described in \cite{Blas:2015tla}.
For the scales relevant for Lyman-$\alpha$ data dark matter interactions are negligible by the time non-linear corrections
start to have an impact. Therefore, we can effectively use the implementation described in \cite{Blas:2013aba}, which allows
for a faster computation time.

\begin{figure}[t]
\begin{center}
\includegraphics[width=0.49\textwidth]{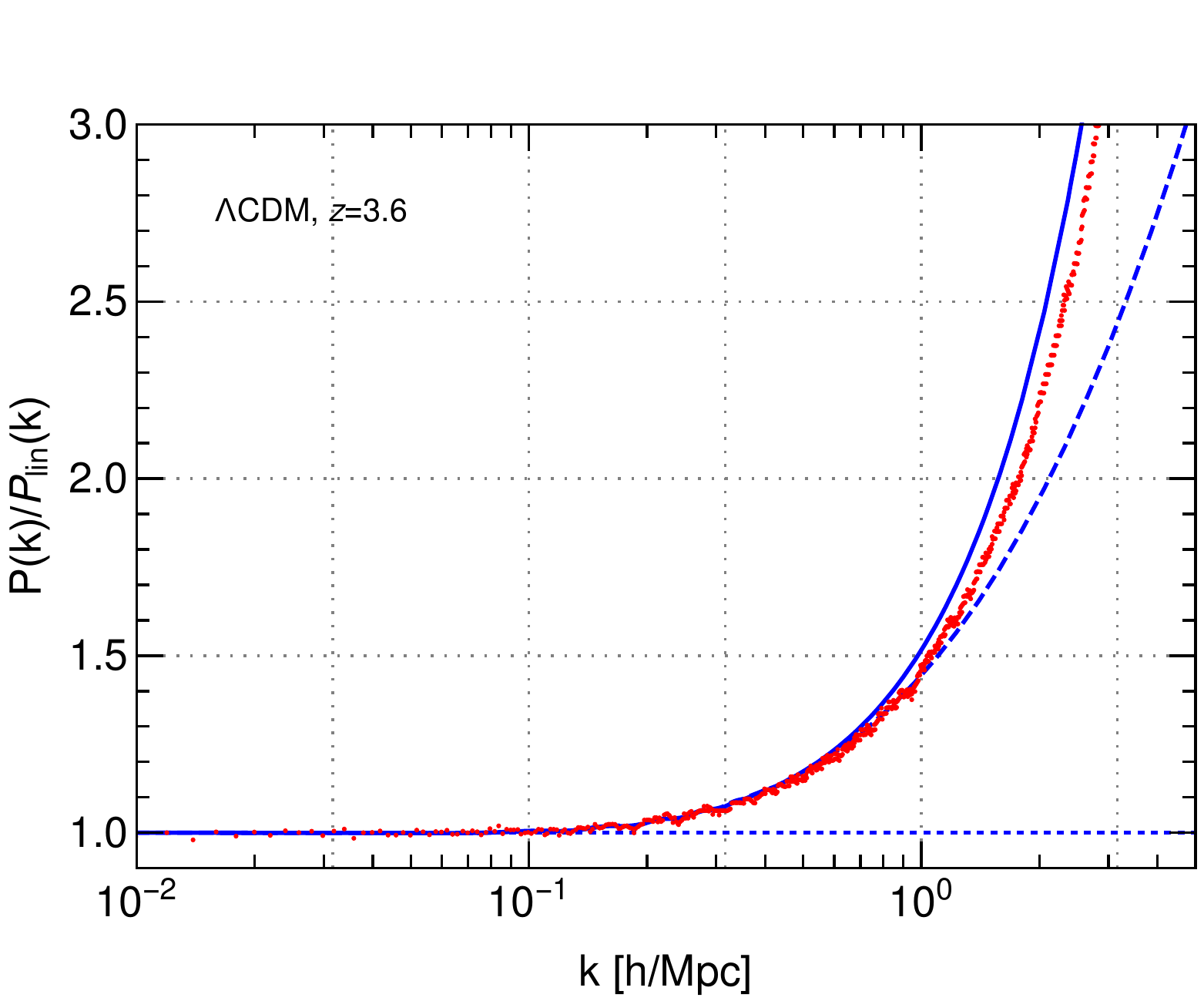}
\end{center}
\caption{\label{fig:hr4}%
\small Comparison of the perturbative matter power spectrum (blue solid 2-loop, dashed 1-loop) with $N$-body data
from the horizon run 4 simulation  \cite{Kim:2015yma} (red points) at $z=3.6$ for $\Lambda$CDM.
}
\end{figure}

As a cross check we consider the matter power spectrum obtained from the horizon run 4 $N$-body simulation \cite{Kim:2015yma}.
In Fig.\,\ref{fig:hr4} we show the matter power spectrum, normalized to the linear spectrum, for $z=3.6$.
The red points show the $N$-body result, the blue solid line corresponds to the two-loop approximation, and the dashed line to one-loop.
Even without adding counterterms the agreement on the relevant scales is at the few percent level for $z=3.6$.

\clearpage

\bibliography{lyalpha}
\bibliographystyle{helsevier}

\end{document}